\documentclass[trackchanges]{aastex701}

\usepackage{comment} 
\usepackage{amsmath}
\usepackage{enumitem}

\begin{document}
\title{Damped Kinetic Alfv\'en Waves in Earth’s Magnetosheath: Numerical Simulations and MMS Observations}

\author[orcid=0009-0000-1368-9263, sname='Chettri']{Mani K Chettri}
\affiliation{Department of Physics, Sikkim University, Gangtok-737102, India}
\email{mkchettri8@gmail.com}

\author[orcid=0009-0003-3061-8944, sname='Singh']{Hemam D. Singh}
\affiliation{Department of Physics, Netaji Subhas University of Technology, New Delhi-110078, India}
\email[show]{hemam.singh@nsut.ac.in}

\author[orcid=0009-0003-6908-7263, sname='Shrivastav']{Vivek Shrivastav}
\affiliation{Department of Physics, Sikkim University, Gangtok-737102, India}
\email{vivekshrivastav1998@gmail.com}

\author[orcid=0009-0009-9813-3402, sname='Singh']{Britan Singh}
\affiliation{Department of Physics, Sikkim University, Gangtok-737102, India}
\email{britansingh@gmail.com}

\author[orcid=0000-0003-3955-7116, sname='Mukherjee']{Rupak Mukherjee}
\affiliation{Department of Physics, Sikkim University, Gangtok-737102, India}
\email{rmukherjee@cus.ac.in}

\begin{abstract}
The Earth's magnetosheath provides a high $\beta$ (ratio of electron thermal pressure to magnetic pressure) plasma environment where kinetic Alfv\'en waves (KAWs) strongly influence turbulence and energy dissipation. This study investigates how Landau damping modifies the nonlinear evolution of KAWs by solving a modified nonlinear Schr\"odinger equation that captures both dispersive and nonlinear effects. Without Landau damping, modulational instability drives rapid self-focusing into intense magnetic filaments, producing a turbulent cascade with $k_\perp^{-5/3}$ scaling in the inertial range ($k_\perp\rho_i<1$) that transitions to $k_\perp^{-8/3}$ at sub-ion scales ($k_\perp\rho_i>1$), here $k_\perp$ is the wavevector component perpendicular to the background magnetic field and $\rho_i$ the ion thermal gyroradius. When Landau damping is included, magnetic structures are significantly suppressed, and the spectrum steepens to $k_\perp^{-11/3}$ in the sub-ion range while the inertial range maintains $k_\perp^{-5/3}$ scaling. The damping acts across all scales through resonant wave-particle interactions, efficiently transferring energy from waves to particles. Direct comparison with Magnetospheric Multiscale (MMS) spacecraft observations shows that the observed kinetic range spectral slope falls between our undamped and damped simulation limits, consistent with an intermediate damping regime in magnetosheath turbulence. This agreement confirms that Landau damping is one of the primary mechanisms controlling turbulent energy dissipation at kinetic scales in collisionless plasmas.
\end{abstract}


\section{Introduction} \label{introduction}
The Earth's magnetosheath, situated just downstream of the bow shock, serves as a dynamic laboratory for space plasma physics. This region is characterized by strong velocity shears, significant pressure gradients, and pronounced fluctuations in electromagnetic fields \citep{rakhmanova2021plasma,pollock2018magnetospheric}. Its accessibility to multi-spacecraft missions such as Magnetospheric Multiscale (MMS), Time History of Events and Macroscale Interactions during Substorms (THEMIS), and Cluster has made it an essential target for observational and theoretical studies. The magnetosheath hosts a rich variety of plasma wave modes and turbulent structures, providing a natural environment to investigate nonlinear processes and energy dissipation in collisionless plasmas \citep{lucek2005magnetosheath,macek2018magnetospheric,rakhmanova2021plasma,stawarz2024interplay}. The bow shock, formed by the interaction of the supersonic solar wind with the Earth's magnetosphere, plays a crucial role in shaping the properties of the magnetosheath plasma. As the solar wind crosses the bow shock, it undergoes a dramatic transformation: the plasma is decelerated, compressed, and heated, leading to a significant increase in density and temperature. This process not only modifies the macroscopic properties of the plasma but also influences microphysical processes, such as the generation of turbulence and the excitation of various wave modes \citep{macek2018magnetospheric}.

In contrast to the solar wind, which is dominated by predominantly outward-propagating Alfv\'en waves, the magnetosheath features a more symmetric distribution of Alfv\'en wave propagation directions. This symmetry fundamentally modifies the turbulence characteristics and the routes through which energy cascades and dissipates \citep{sibeck2007alfven,alexandrova2008solar,rakhmanova2019turbulent}. The combination of these Alfv\'enic fluctuations with high $\beta$ plasma supports the coexistence of fluid and kinetic effects, fostering multiscale energy transfer and dissipation \citep{schekochihin2009astrophysical,chen2017nature}, where $\beta$ ($=8\pi n_0k_BT_e/B_0^2$) is the ratio of the electron thermal pressure to the magnetic pressure, $n_0$ is the equilibrium plasma density, $T_e$ is the electron temperature and $B_0$ is the background (ambient) magnetic field. In the Earth's magnetosheath, the plasma $\beta$ typically ranges from 1 to 10 \citep{lucek2005magnetosheath,alexandrova2008spectra}. This range reflects the dynamic interaction between the thermal energy of the plasma and the energy stored in the magnetic field \citep{russell1990upstream}. Higher plasma $\beta$ values indicates that thermal pressure dominates \citep{lu2015pressure,alexandrova2008spectra}, which is characteristic of the magnetosheath \citep{ma2020statistical} where the plasma is heated and compressed by the bow shock\citep{breuillard2018new,macek2018magnetospheric}.

At large scales, plasma fluctuations within the magnetosheath closely resemble magnetohydrodynamic (MHD) Alfv\'en waves. However, as the cascade reaches ion kinetic scales, specifically when $k_\perp \rho_i \sim 1$, the dominant dynamics shift to kinetic Alfv\'en waves (KAWs) \citep{howes2014kinetic,chen2017nature}, here, \(k_\perp\) is the wavevector component perpendicular to the ambient magnetic field and $\rho_i$ is the ion thermal gyroradius. By analyzing recent MMS observations using burst-mode magnetic field data with high-resolution of 7.8 ms, \citet{macek2018magnetospheric} were able to quantify this transition with considerable precision. They found clear spectral breaks near the ion gyrofrequency (around 0.25\,Hz), occurring systematically across different magnetosheath locations with plasma $\beta$ values of 2.2--4.9, with magnetic energy spectra steepening from approximately $-0.8$ to $-5/2$ above the ion gyrofrequency. At even higher frequencies (above roughly 20 Hz), the spectra become steeper still, showing slopes ranging from $-7/2$ to $-11/2$ (some reaching $-16/3$). The observed spectral steepening to $\sim k_\perp^{-3.2}$ in the dissipation range \citep{wilder2018role}, combined with the systematic occurrence of parallel electric fields during energy conversion processes, provides what we consider direct evidence of KAW dissipation processes. What makes these findings particularly compelling is that these spectral signatures appear systematically across different magnetosheath locations. The observational picture is further supported by hybrid Vlasov--Maxwell simulations, which confirm the dominance of KAWs and highlight the critical role that dissipation mechanisms play at ion kinetic scales \citep{perrone2018fluid}.

How turbulent energy ultimately transforms into plasma heating remains one of the key unresolved problems in astrophysical plasmas. While the complete picture is not yet clear, collisionless damping through wave-particle interactions is thought to be one of the leading mechanisms \citep{tenbarge2013collisionless}. This process is particularly relevant in magnetosheath plasmas ($\beta \sim 1$--$10$), where the $\beta$-dependent dispersion properties of KAWs naturally bring their phase velocity into resonance with the electron thermal distribution as demonstrated in both spacecraft observations and kinetic simulations by \citet{afshari2021importance} and \citet{horvath2020electron}. Such resonant energy transfer drives significant electron heating that has been directly observed through spacecraft measurements \citep{chen2019evidence}. Although classical Landau damping theory is well established for electrostatic waves, understanding its role in electromagnetic KAWs, particularly under nonlinear turbulent conditions, remains an active research area.

These KAWs constitute a dispersive extension of the shear Alfv\'en mode, governed by finite ion gyroradius and electron inertia effects \citep{goertz1979magnetosphere,sharma2024kinetic}. Crucially, KAWs exhibit two defining characteristics: (i) nonzero parallel electric fields and (ii) perpendicular magnetic compressibility \citep{hasegawa1976particle,hollweg1999kinetic}. These properties favor resonant wave-particle interactions with electrons satisfying $v_\parallel \approx \omega/k_\parallel$ \citep{hollweg1999kinetic,gershman2017wave,rakhmanova2019turbulent}, giving rise to Landau damping which is a key collisionless dissipation mechanism in high $\beta$ plasma environments such as the magnetosheath \citep{horvath2020electron,howes2014kinetic}, where $v_\parallel$ is the electron velocity component parallel to the ambient magnetic field ($\mathbf{B_0}$) direction, $\omega$ is the frequency of the pump KAW and $k_\parallel$ is the wave vector component parallel to $\mathbf{B_0}$. Throughout this paper, 'parallel' and 'perpendicular' refer to directions relative to the background magnetic field, $\mathbf{B_0}$.

Various spacecraft analyses have reported signatures of KAWs in magnetosheath reconnection events \citep{rakhmanova2021plasma,stawarz2022turbulence,roberts2022kinetic}, with direct MMS observations revealing nearly monochromatic KAW-branch wave packets \citep{gershman2017wave} and KAWs exhibiting dominant frequencies at $\sim$0.38--0.64~Hz with perpendicular wavelengths comparable to the ion thermal gyroradius \citep{teh2023kinetic}. Using field-particle correlation applied to MMS spacecraft data, \citet{chen2019evidence} presented direct measurements of electron energy transfer linked to KAW turbulence, revealing characteristic velocity-space signatures of Landau damping that substantiate earlier theoretical predictions. A comprehensive analysis by \citet{afshari2021importance} of 20 MMS magnetosheath intervals revealed that Landau damping is present in 95\% of the cases, with velocity-space signatures consistent with linear and nonlinear kinetic theory for KAWs.

Despite this clear observational evidence of ubiquitous Landau damping, accurately modeling these nonlinear kinetic processes in turbulence simulations remains challenging. Fluid/hybrid simulations often neglect important kinetic resonances, while fully kinetic models are computationally prohibitive for large-scale turbulence. Consequently, predictions of energy partition (e.g., ion temperature anisotropy) and spectral breaks near \( k_\perp \rho_i \sim 1 \) in fluid models diverge from kinetic simulations. Specifically, fluid approaches generally overestimate compressibility and underrepresent kinetic dissipation mechanisms like Landau damping \citep{perrone2018fluid}. Most numerical models of KAW dynamics in space plasmas still rely on fluid or hybrid frameworks that either neglect Landau damping entirely or incorporate it only implicitly, limiting their applicability to resonant dissipation processes.

Previously, \cite{chettri2024nonlinear} developed a two-fluid model of nonlinear coupling of KAWs and ion acoustic waves generated by the ponderomotive force of the KAWs in nonadiabatic regime, applicable to an arbitrary beta plasma environment. It was numerically studied to understand some physical processes in the Sun's near streamer belt and the Earth's radiation belt regions that captured key phenomena such as the formation of magnetic field structures and their collapse, the evolution from periodic to chaotic density perturbations, Kolmogorov-like inertial-range spectral cascades with a $-5/3$ index, and steeper dissipation-range spectra approaching $-3$ to $-4$, along with indications of stochastic particle heating leading to superthermal tails in distribution functions. However, this model did not self-consistently account for the energy transfer from KAWs to particles via Landau damping, particularly under the high $\beta$ plasma conditions characteristic of the magnetosheath where kinetic effects dominate wave-particle interactions. To address this gap, we extend our two-fluid model by including a Landau damping phenomenologically in the equations of motions. The KAW dynamics is coupled nonlinearly with the magnetosonic waves (MSW) through the density modified by the KAW ponderomotive force. The dynamical equation comes out of the form of modified nonlinear Schr\"odinger equation (NLSE), which is solved using Fourier pseudospectral methods for high spectral accuracy (with de-aliasing). The simulation results are then systematically compared with MMS in-situ measurements to assess their validity. 

The structure of this paper is as follows. In Section~\eqref{sec:model_equations}, we derive the theoretical model in the form of modified nonlinear Schr\"{o}dinger equation governing KAW dynamics with Landau damping. Section~\eqref{sec:numerical_simulation} outlines the numerical methodology. In Section~\eqref{sec:results}, we present the simulation results and a discussion, and in Section~\eqref{sec:comparision} we compare these results with \emph{MMS} spacecraft observations. Finally, Section~\eqref{conclusion} summarizes our findings and their implications for turbulent dissipation in high $\beta$ magnetosheath plasmas.


\section{Model Equations}
\label{sec:model_equations}
We take a collisionless, non‐relativistic two‐fluid plasma (electrons and protons) with a constant magnetic field $\mathbf{B}_0 = B_0\,\hat z$ and an electric field of small fluctuations, $\mathbf{E} = \delta\mathbf{E}$. A pump KAW is propagating in the $x$–$z$ plane, with wavevector $\mathbf{k}_0 = k_{0x}\,\hat x + k_{0z}\,\hat z$, so that all variations occur in $x$ and $z$, except for the magnetic perturbation $\delta B_y$ in the $y$ direction. The two-fluid description captures the essential KAW physics through electron inertia, parallel electron pressure gradients, and the decoupling of ion and electron motions. While full finite-Larmor-radius corrections become crucial for $k_\perp \rho_i > 1$, in the present regime the dominant effects arise from electron kinetics and the breakdown of ideal MHD approximations. Electrons remain strongly magnetized and stream rapidly along $B_0$, making their parallel motion and pressure gradients essential for KAW dynamics \citep{miloshevich2021inverse}. The model naturally incorporates these electron kinetic effects while maintaining a fluid description of both species

The linearized continuity and momentum equations are, respectively, given by
\begin{equation}\label{continuity1}
\partial_t \delta n_j + n_{0j} \,\nabla\!\cdot\!\delta\mathbf{v}_j \approx 0
\end{equation}
and
\begin{equation}
\label{eq:momentum}
m_j\left(\partial_t \delta \mathbf{v}_j + \gamma_L \delta \mathbf{v}_j\right) \approx q_j \delta \mathbf{E} + \frac{q_j}{c}(\delta \mathbf{v}_j \times \mathbf{B}_0) - \frac{\gamma_j k_B T_j}{n_{0j}} \nabla \delta n_j,
\end{equation}
where the subscript $j$ denotes the particle species (electrons $e$ and ions $i$). Here, $m_j$, $q_j$, $T_j$, and $n_{0j}$ represent the mass, charge, temperature, and equilibrium density of species $j$, while $\delta n_j$ and $\delta\mathbf{v}_j$ denote the perturbed density and bulk velocity, respectively. The quantities $\delta\mathbf{E}$ and $\mathbf{B}_0$ are the perturbed electric field and background magnetic field, $c$ is the speed of light, and $k_B$ is the Boltzmann constant. The coefficient $\gamma_L$ represents a phenomenological Landau damping rate, and $\gamma_j$ is the ratio of specific heats ($c_p/c_v$). We adopt the shorthand notation $\partial_t \equiv \partial/\partial t$ and $\partial_x \equiv \partial/\partial x$ (and similarly for other derivatives). The system is assumed to be quasineutral ($n_{0e} \simeq n_{0i} \simeq n_0$) and isothermal ($\gamma_e = \gamma_i = 1$), with small perturbations ($\delta n_j/n_{0j} \ll 1$). The nonlinear convective term $(\mathbf{v}\!\cdot\!\nabla)\mathbf{v}$ is neglected; this approximation is justified under weak dispersion and remains valid for perpendicular motion even in the strong dispersion limit \citep{kaur2016ion,sadiq2018linear}.

In our two-fluid model, we introduce Landau damping through a phenomenological damping rate $\gamma_L$; its expression in terms of wave number is given in the last part of this section (Equation \ref{eq:gamma_L}) . While true Landau damping is a kinetic process arising from resonant wave-particle interactions, our simplified treatment provides a first-order approximation of dissipative effects within the fluid framework. This approach allows us to isolate the general impact of collisionless energy dissipation on the nonlinear evolution of KAW turbulence, providing a valuable contrast to the non-dissipative case.

Using Eqs.~\eqref{continuity1} and \eqref{eq:momentum} and assuming all the first order fluctuations to be of the form of \(\delta f=\delta \tilde{f} e^{i\left(k_{0x}x+k_{0z}z -\omega t\right)}\), we can express the component-wise electron and ion velocities respectively, as follows:
\begin{align}
\delta v_{ex} &= -\frac{e}{m_e} \frac{ (\gamma_L - i\omega) \delta E_x - \omega_{ce} \delta E_y }{ (\gamma_L - i\omega)^2 + \omega_{ce}^2 } - \frac{k_B T_e}{m_e n_0} \frac{ \gamma_L - i\omega }{ (\gamma_L - i\omega)^2 + \omega_{ce}^2 } \partial_x \delta n,
\label{eq:evx_sol}
\end{align}

\begin{align}
\delta v_{ey} &= -\frac{e}{m_e} \frac{ (\gamma_L - i\omega) \delta E_y + \omega_{ce} \delta E_x }{ (\gamma_L - i\omega)^2 + \omega_{ce}^2 } - \frac{k_B T_e}{m_e n_0} \frac{ \omega_{ce} }{ (\gamma_L - i\omega)^2 + \omega_{ce}^2 } \partial_x \delta n,
\label{eq:evy_sol}
\end{align}

\begin{equation}
\delta v_{ez} = -\frac{e\,\delta E_z}{m_e(\gamma_L - i\omega)} - \frac{k_B T_e}{m_e n_0(\gamma_L - i\omega)}\,\partial_z\delta n,
\label{eq:evz_sol}
\end{equation}

\begin{equation}
\delta v_{ix} = \frac{e}{m_i} \frac{ (\gamma_L - i\omega) \delta E_x + \omega_{ci} \delta E_y }{ (\gamma_L - i\omega)^2 + \omega_{ci}^2 } 
- \frac{k_B T_i}{m_i n_0} \frac{ \gamma_L - i\omega }{ (\gamma_L - i\omega)^2 + \omega_{ci}^2 } \partial_x \delta n,
\label{eq:ivx_sol}
\end{equation}

\begin{equation}
\delta v_{iy} = \frac{e}{m_i} \frac{ (\gamma_L - i\omega) \delta E_y - \omega_{ci} \delta E_x }{ (\gamma_L - i\omega)^2 + \omega_{ci}^2 } 
+ \frac{k_B T_i}{m_i n_0} \frac{ \omega_{ci} }{ (\gamma_L - i\omega)^2 + \omega_{ci}^2 } \partial_x \delta n
\label{eq:ivy_sol}
\end{equation}
~~ and
\begin{equation}
\delta v_{iz} = \frac{e \delta E_z}{m_i (\gamma_L - i\omega)} 
- \frac{k_B T_i}{m_i n_0 (\gamma_L - i\omega)} \partial_z \delta n,
\label{eq:ivz_sol}
\end{equation}
where \(\omega_{cj}=q_jB_0/m_jc\) is the cyclotron frequency of species $j$.

Considering the low-frequency ordering (\(\omega\ll\omega_{ci}\)) and differentiating Eq.~\eqref{eq:ivx_sol} with respect to \(x\), we obtain
\begin{equation}
\partial_x\delta v_{ix}
\approx \frac{e}{m_i\omega_{ci}^2}\Bigl[\omega_{ci}\,\partial_x\delta E_y + (\gamma_L - i\omega)\,\partial_x\delta E_x\Bigr]
- \frac{k_B T_i}{m_i n_0}\frac{\gamma_L - i\omega}{\omega_{ci}^2}\,\partial_{xx} \delta n.
\label{eq:dvixdx}
\end{equation}

The linearized electron and ion continuity equations can be respectively written as
\begin{equation}
\partial_t \delta n + n_0 \partial_z \delta v_{ez} \approx 0,
\label{eq:lin_continuity_e}
\end{equation}
and 
\begin{equation}
\partial_t {\delta n} + n_{0}\partial_x \delta{v_{ix}} = 0. 
\label{eq:lin_continuity_i}
\end{equation}

In writing Equation \eqref{eq:lin_continuity_e} we drop the term $\partial_x\delta v_{ex}$. In the KAW regime ($\omega\ll\omega_{ce}$), electrons are strongly tied to the magnetic field and move primarily along $\mathbf{B}_0$; their small gyroradius and low mass suppress perpendicular (polarization) drifts, so the perpendicular electron velocity is negligible and parallel motion dominates. We also neglect parallel ion motion in Equation ~\eqref{eq:lin_continuity_i} because ions' large mass and Larmor radius make their response predominantly perpendicular. Using plane waves ($\partial_t\to -i\omega$, $\partial_z\to ik_{0z}$), Equation \eqref{eq:lin_continuity_e} is transformed to
\begin{equation}
\delta n = \frac{n_0 k_{0z}}{\omega}\,\delta v_{ez}.
\label{eq:ne_vez}
\end{equation}

To obtain the dynamical equation for the transverse magnetic perturbation $\delta B_y$ of the pump KAWs, we begin by expressing it in terms of the perpendicular and parallel electric field perturbations, $\delta E_x$ and $\delta E_z$. From Faraday’s law,
\begin{equation}
    \nabla \times \mathbf{E} = -\frac{1}{c}\,\partial_t \mathbf{B},
    \label{eq:farday}
\end{equation}
taking the $y$-component, followed by a time derivative, gives the relation between the magnetic and electric field perturbations as:
\begin{equation}
    \partial_{tt}\delta B_y 
    = c \left( \partial_{tx}\delta E_z - \partial_{tz}\delta E_x \right).
    \label{eq:Faraday_by}
\end{equation}

Substituting Equation \eqref{eq:evz_sol} into Equation \eqref{eq:ne_vez} to relate $\delta v_{ez}$ and $\delta E_z$, we get
\begin{equation}
\delta v_{ez}\left[1 + \frac{ik_{0z}^2 k_B T_e}{m_e \omega(\gamma_L - i\omega)}\right]
= -\frac{e\,\delta E_z}{m_e(\gamma_L - i\omega)}.
\end{equation}

Under the low-frequency approximation \(\omega\ll\omega_{pe}\), the displacement current may be neglected since the wave phase velocity \(v_{\rm ph}\) is very small than the speed of light. If we denote the characteristic temporal and spatial scales of the perturbations by \(\tau\) and \(l\), respectively, and estimating \(\partial/\partial t\sim\tau^{-1}\) and \(\nabla\sim l^{-1}\) gives \(v_{\rm ph}\sim l/\tau\). From Faraday's law we can write \(E/B\sim v_{\rm ph}/c\). Hence, the ratio of the displacement current \((1/c)\,\partial_t\mathbf{E}\) to the conduction current \(\nabla\times\mathbf{B}\) scales like \((v_{\rm ph}/c)^2\ll1\) and can be dropped. Now, Amp\'ere's law can be written as 
\begin{equation}
\nabla\times\delta\mathbf{B}=\frac{4\pi}{c}\,\delta\mathbf{J},
\label{eq:ampere}
\end{equation}
where the perturbed current density is
\begin{equation*}
\delta\mathbf{J}=e n_0\big(\delta\mathbf{v}_i-\delta\mathbf{v}_e\big).
\end{equation*}

Taking the \(z\)-component of \eqref{eq:ampere} and differentiating with respect to time yields
\begin{equation}
\partial_{t x}\,\delta B_y=\frac{4\pi}{c}\,\partial_t\delta J_z,
\label{eq:by_t_x}
\end{equation}
where \(\delta J_z\) denotes the field-aligned (parallel) component of the perturbed current density. 

Subtracting Equations~\eqref{eq:lin_continuity_e} and \eqref{eq:lin_continuity_i} and imposing quasi-neutrality ($n_i \simeq n_e$) correctly yields the local conservation of current, $\nabla \cdot \mathbf{J} = 0$. In KAWs, the parallel current is carried predominantly by electrons, as ion parallel motion is negligible. Thus, the parallel current density becomes
\begin{equation}
\delta J_z \simeq -e n_0 \delta v_{ez}.
\label{eq:J_z}
\end{equation}
The field-aligned current density $\delta J_z$ thus generated produces the transverse magnetic perturbation $\delta B_y$ through Amp\'ere's law and will be used to couple electron dynamics to the magnetic response of the KAW.

Using the linearized parallel current density from Equation~\eqref{eq:J_z}, Equation~\eqref{eq:by_t_x} can be transformed by substituting the expression for $\delta v_{ez}$. The relation then becomes
\begin{equation}
\partial_{tx} \delta B_y
= -\frac{\omega_{pe}^2\, i\omega}{c(\gamma_L - i\omega)}
\Biggl[\delta E_z + \frac{k_B T_e}{e n_0}\,\partial_z \delta n\Biggr],
\label{eq:by_expr}
\end{equation}
where $\omega_{pe}$ ($=\sqrt{4\pi n_0 e^2 / m_e}$) is the electron plasma frequency.

Solving for the parallel component of the electric field ($\delta E_z$) from the linearized Equation~\eqref{eq:by_expr}, we get
\begin{equation}
\delta E_z
= -\frac{c(\gamma_L-i\omega)}{\omega_{pe}^2 i\omega}\,\partial_{tx}\delta B_y
- \frac{k_B T_e}{e n_0}\,\partial_z\delta n.
\label{eq:Ez_from_By}
\end{equation}

In writing the conservation law of current density, $\nabla \cdot \mathbf{J} = 0$, we use the KAW approximations that perpendicular current is carried by ions and parallel current is carried by electrons. This gives the relation in terms of velocity components as
\begin{equation}\label{eq:J_z2}
\partial_x \left(\delta{v_{ix}}\right) - \partial_z \left(\delta{v_{ez}}\right) = 0.
\end{equation}
Using Equation~\eqref{eq:dvixdx} for $\partial_x \delta v_{ix}$ and the relation for \(\partial_z \delta v_{ez} = -[ec/(m_e \omega_{pe}^2)]\partial_{xz} \delta B_y\) (derived from Amp\'ere's law after accounting for the plane wave assumption), we substitute into Equation~\eqref{eq:J_z2}. Neglecting the ion pressure term (effectively assuming cold ions, $T_i \approx 0$) and keeping the leading linear terms, we obtain the following relation:
\begin{equation}
(\gamma_L - i\omega)\,\partial_x\delta E_x
=  - \omega_{ci}\,\partial_x\delta E_y
- \frac{m_i c\,\omega_{ci}^2}{m_e\omega_{pe}^2}\,\partial_{xz}\delta B_y.
\label{eq:Ex_relation}
\end{equation}

The $z$-component of Faraday's law (with $\partial_y = 0$) gives
\begin{equation}
\delta E_y=\frac{\omega}{c k_{0x}}\,\delta B_z.
\label{eq:Ey_Bz}
\end{equation}

Under low-frequency conditions in low-\(\beta\) plasmas, the compressive magnetic perturbation \(\delta B_z\) typically remains negligible \citep{howes2006astrophysical, schekochihin2009astrophysical, cramer2011physics}. However, in the case of an arbitrary finite $\beta$ plasma, $\delta B_z$ becomes significant and can be calculated from the pressure balance condition \(\nabla\left(k_B T\delta n+ \delta B_z^2/8\pi\right)=0\). This leads us to the relation $\delta B_z/B_0 = -\beta \delta n_e/2n_0$, which reveals a strong anti-correlation between magnetic and thermal pressures, consistent with what \citet{hollweg1999kinetic} found in his work. We can use this relationship to investigate the density and magnetic field fluctuations that appear in the inertial range of magnetic turbulence spectra \citep{burlaga1990pressure,roberts1990heliocentric}. Now, if we combine the continuity equation (\ref{continuity1}) with Amp\'ere's law \eqref{eq:ampere}, we obtain
\begin{equation}
\delta B_z=-\frac{\beta}{2}\,\frac{\omega_{ce}c^2}{\omega_{pe}^2}\,
\frac{k_{0x}}{\omega}\,\partial_z(\delta B_y),
\label{eq:Bz_from_By}
\end{equation}
We can then rearrange Equation \eqref{eq:Bz_from_By} to get
\begin{equation}
     \frac{\delta B_z}{\delta B_y} = -\frac{ik_{0x}k_{0z}c_s^2}{\omega \omega_{ci}},
     \label{eq:Bz_By}
\end{equation}
which matches Equation (11) from \citet{hollweg1999kinetic}.

Substituting Equation~\eqref{eq:Bz_from_By} into Equation~\eqref{eq:Ey_Bz}, we obtain $\delta E_y$ as
\begin{equation}
\delta E_y = -\frac{\beta}{2}\,\frac{\omega_{ce}\,c}{\omega_{pe}^2}\,\partial_z \delta B_y.
\label{eq:E_y}
\end{equation}
Using Equation~\eqref{eq:E_y} in Equation~\eqref{eq:Ex_relation} gives the expression for $\delta E_x$.
\begin{equation}
(\gamma_L - i\omega)\, \delta E_x
= \frac{\beta}{2}\,\frac{\omega_{ci}\,\omega_{ce}\,c}{\omega_{pe}^2}\,\partial_z (\delta B_y)
- \frac{m_i\,c\,\omega_{ci}^2}{m_e\,\omega_{pe}^2}\,\partial_z (\delta B_y). 
\label{eq:E_x}
\end{equation}

Upon substituting the expressions for the electric field components, Equations~\eqref{eq:E_x} and \eqref{eq:Ez_from_By}, into Faraday's law, Equation~\eqref{eq:Faraday_by}, after a careful simplification, we obtained the dynamical equation for the weakly damped nonlinear KAW:
\begin{align}
&\partial_{tt}\delta B_y + 2\gamma_L \partial_t \delta B_y - \lambda_e^2 \partial_{ttxx}\delta B_y - 2\gamma_L \lambda_e^2 \partial_{txx}\delta B_y
\nonumber\\
&\qquad
- v_A^2\left(1-\frac{\delta n}{n_0}\right)\partial_{zz}\delta B_y + v_{te}^2 \lambda_e^2 \partial_{xxzz}\delta B_y = 0.
\label{eq:kaw_modified}
\end{align}
Here, $\lambda_e = c/\omega_{pe}$ is the electron inertial length, $v_{te} = \sqrt{k_B T_e / m_e}$ is the electron thermal speed, and $v_A = B_0 / \sqrt{4\pi n_0 m_i}$ is the Alfv\'en speed. A representative solution to Equation \eqref{eq:kaw_modified} is a linearly polarized plane wave at a fundamental frequency $\omega$, modulated by a slowly varying envelope, expressed as
\begin{equation}
\delta{B_y} = \delta \Tilde{B_y}(x,z,t) \, \exp\left[i(k_{0x}x + k_{0z}z - \omega t)\right],
\label{eq:envelope_by}
\end{equation}
where $\delta \Tilde{B_y}(x,z,t)$ denotes the envelope function, which describes the inhomogeneous amplitude of the transverse magnetic field. This envelope varies slowly in space compared to the rapid phase oscillations of the carrier wave, $\exp\left[i(k_{0x}x + k_{0z}z - \omega t)\right]$. By taking the Fourier transform to the linear part of Equation \eqref{eq:kaw_modified}, we get the linear dispersion relation of KAW as
\begin{equation}
\big(1+\lambda_e^2 k_{0x}^2\big)\,\omega^2 + 2i\gamma_L\big(1+\lambda_e^2 k_{0x}^2\big)\,\omega - k_{0z}^2\Big(v_{te}^2\lambda_e^2 k_{0x}^2 + v_A^2\Big)=0.
\label{eq:disp_kaw_new}
\end{equation}
Solving in the weak damping limit ($\gamma_L \ll \omega_0$) and retaining first-order terms in $\gamma_L$, the dispersion relation becomes
\begin{equation}
\frac{\omega^2}{k_{0z}^2} = v_A^2\frac{1 + \rho_s^2 k_{0x}^2}{1+\lambda_e^2 k_{0x}^2} - \frac{2i\gamma_L v_A}{k_{0z}} \sqrt{\frac{1 + \rho_s^2 k_{0x}^2}{1+\lambda_e^2 k_{0x}^2}},
\label{eq:disp_kaw_explicit}
\end{equation}
where $\rho_s = c_s / \omega_{ci}$ is the ion sound gyroradius, expressed as $\rho_s^2 = (\beta/2)\lambda_e^2$ with $\beta = 2v_{te}^2/v_A^2$. This form indicates that the dispersion arises from the competition between electron inertia (denominator) and electron thermal pressure (numerator term $\rho_s^2 k_{0x}^2$). In the long wavelength limit ($k_{0x}\rho_s \ll 1$, $k_{0x}\lambda_e \ll 1$), the relation reduces to $\omega^2 \approx k_{0z}^2 v_A^2$, recovering the shear Alfv\'en wave. At intermediate perpendicular scales where $k_{0x}\rho_s \gg 1$ but $k_{0x}\lambda_e \lesssim 1$, the dispersion exhibits the characteristic KAW scaling $\omega^2/k_{0z}^2 \sim v_A^2 \rho_s^2 k_{0x}^2$. In the regime where both $k_{0x}\rho_s \gg 1$ and $k_{0x}\lambda_e \gg 1$, the parallel phase velocity saturates at the electron thermal speed, $\omega/k_{0z} \sim v_{te}$. It is worth noting that in this cold-ion fluid model, $\rho_s$ serves as the effective dispersive scale, playing the role typically assigned to the ion thermal gyroradius $\rho_i$ in fully kinetic theories. Thus, the model is valid from MHD scales ($k_\perp\rho_s \ll 1$) through the KAW regime ($k_\perp \rho_s \sim 1$), but implies limitations at very short wavelengths where fully kinetic effects dominate.

The envelope equation form of the dynamical equation for the evolution of KAW can be obtained by utilizing Equation \eqref{eq:envelope_by} into Equation \eqref{eq:kaw_modified} as: 
\begin{eqnarray}
&&
2\left(\gamma_L - i\omega - i\lambda_e^2\omega k_{0x}^2 + \gamma_L\lambda_e^2 k_{0x}^2\right) \partial_t \delta \tilde{B}_y
+ 2\left(i\omega^2\lambda_e^2 k_{0x} - 2\gamma_L\lambda_e^2\omega k_{0x} - ik_{0x} k_{0z}^2 v_{te}^2\lambda_e^2\right) \partial_x \delta\tilde{B}_y
\nonumber \\ &&
+ \left(\omega^2\lambda_e^2 + 2i\omega\gamma_L\lambda_e^2 - v_{te}^2 \lambda_e^2 k_{0z}^2\right) \partial_{xx} \delta\tilde{B}_y
+ 2\bigg[\frac{ik_{0z} B_0 c\beta \omega_{ci}}{8\pi n_0 e}
- ik_{0x}^2 k_{0z} v_{te}^2\lambda_e^2
- k_{0z} v_A^2\Bigl(1 - \frac{\delta n}{n_0}\Bigr) \bigg]
\partial_z \delta \tilde{B}_y
\nonumber \\ &&
- \bigg[ k_{0x}^2 v_{te}^2\lambda_e^2 - \frac{B_0 c\beta \omega_{ci}}{8\pi n_0 e}
+ v_A^2 \Bigl(1 - \frac{\delta n}{n_0}\Bigl) \bigg]
\partial_{zz} \delta \tilde{B}_y - 4\left(\omega k_{0x} \lambda_e^2 + i\gamma_L\lambda_e^2 k_{0x}\right) \partial_{tx} \delta \tilde{B}_y
\nonumber \\ &&
+ 2\left(i\omega\lambda_e^2 - \gamma_L\lambda_e^2\right) \partial_{txx} \delta \tilde{B}_y
- 4 k_{0x} k_{0z} v_{te}^2\lambda_e^2 \partial_{xz} \delta \tilde{B}_y
+ 2i v_{te}^2 \lambda_e^2 k_{0z} \partial_{xxz} \delta \tilde{B}_y
+ v_{te}^2\lambda_e^2 \partial_{xxzz} \delta \tilde{B}_y
\nonumber \\ &&
+ \bigg[ v_A^2k_{0z}^2\Bigl(1 - \frac{\delta n}{n_0}\Bigl)
- \frac{B_0 c\beta \omega_{ci} k_{0z}^2}{8\pi n_0 e} + k_{0x}^2 k_{0z}^2 v_{te}^2\lambda_e^2
- 2i\omega\gamma_L\lambda_e^2 k_{0x}^2 - \lambda_e^2\omega^2 k_{0x}^2 - \omega^2 - 2i\omega\gamma_L \bigg] \delta \tilde{B}_y = 0
\label{eq:kaw_envelope}
\end{eqnarray}

The total density perturbation ($\delta n/n_0$) in Equation \eqref{eq:kaw_envelope} arises from the coupling of the KAW with other plasma modes \citep{shukla1999plasma}. In our study, we have taken the density perturbation arising from the two physical contributions:
\begin{equation}\label{eq:density}
    \frac{\delta n}{n_0}= \frac{\delta n_1}{n_0} + \frac{\delta n_2}{n_0}
\end{equation}

The first component, $\delta n_1 / n_0$, results from the nonlinear coupling between the KAW and a MSW driven by the ponderomotive force of the pump KAW. We take the second term as $\delta n_2 / n_0 = \eta \cos(\alpha_z z)$ representing a large-scale density modulation along the magnetic field direction. Such density variations are commonly observed in turbulent space plasmas like the magnetosheath, where they appear as quasi-static striations or filamentary structures. We use a simple sinusoidal form to capture this background inhomogeneity, where $\eta$ represents the relative modulation amplitude (set to $0.1$, consistent with typical magnetosheath density fluctuations) and $\alpha_z$ is the characteristic wavenumber along the field. The key physical requirement for our analysis is $\alpha_z \ll k_{0z}$, which ensures that the background density varies slowly compared to the KAW wavelength. Under this condition, the KAW envelope responds adiabatically to the density gradient, allowing us to treat the wave as propagating through a gradually varying medium rather than scattering off sharp density jumps. This separation of scales between the background inhomogeneity and the pump wave is essential for the validity of our envelope modulation approach.

Here, we are deriving the density fluctuations ($\delta n_1/n_0$) arising from the dynamics of MSW under the influence of ponderomotive force driven by the pump KAW. Let us consider a MSW wave propagating along the $x$-direction ($\mathbf{k}=k_{1x}\hat{x}$) and polarized along the $y$-direction, $\mathbf{E_1}=\delta E_{1y}\hat{y}$ in the magnetized plasma with magnetic field \( \mathbf{B}=B_0\hat{z} + \delta B_y \hat{y}\). The linearized continuity and momentum equations are respectively written as:
\begin{equation}\label{IAW cont}
    \partial_t \left( \delta n_j \right) + n_{0j}\partial_x\left( \delta v_{1jx}\right) =0
\end{equation}
\begin{equation}\label{IAW main}
    m_j\Bigl(\partial_t \left(\delta \mathbf{v_{1j}}\right) + \delta v_{1jx}\partial_x \left(\delta \mathbf{v_{1j}}\right)\Bigr) = q_j\left(\delta \mathbf{E_1}+\frac{\delta \mathbf{v_{1j}}\times\left(B_0\hat{z}+\delta B_y\hat{y}\right)}{c}\right) - \frac{k_BT_j}{n_{0j}}\partial_x \left(\delta n_j\right)\hat{x}.
\end{equation}
Here, we apply first-order linearization to the continuity equation, while the momentum equation requires both first-order and second-order linearization. We use the second-order terms to capture the MSW dynamics driven by ponderomotive effects of the pump KAWs. The ponderomotive force is given by
\begin{equation}
    \mathbf{F}_j= \frac{q_j}{c}\left(\delta \mathbf{v}_{1j}\times\delta B_y\hat{y}\right)-m_j\delta v_{1jx}\partial_x \left(\delta \mathbf{v}_{1j}\right) ,
\end{equation}
where the first term represents the Lorentz force and the second term is the convective nonlinearity.

Under the low-frequency ordering $\Omega \ll \omega_{cj}$ and quasi-neutrality ($\delta n_e \approx \delta n_i \equiv \delta n_1$), where $\Omega$ is the characteristic frequency of the MSW, and retaining only the dominant ponderomotive contribution and neglecting higher-order ponderomotive drifts which scale as $(\Omega/\omega_{cj})^2$, the $y$-component momentum equations for electrons and ions include the polarization, diamagnetic, and ponderomotive drifts, can be written as:
\begin{align}
 \delta v_{1ey} &= -\frac{i\Omega}{\omega_{ce}^2} \frac{e\delta E_{1y}}{m_e} + \frac{n_0 ck_B T_e}{e B_0 n_0} \partial_x (\delta n_1) + \frac{c F_{ex}}{e B_0},\label{eq:msw_ey}\\
 \delta v_{1iy} &= \frac{i\Omega}{\omega_{ci}^2} \frac{e\delta E_{1y}}{m_i} - \frac{n_0c k_B T_i}{e B_0 n_0} \partial_x (\delta n_1) - \frac{c F_{ix}}{e B_0},\label{eq:msw_iy}
\end{align}
where $F_{ex}$ and $F_{ix}$ are the $x$-components of the nonlinear ponderomotive force acting on electrons ($e$) and ions ($i$). Here, the first term represents the polarization drift ($\delta v_p \propto \partial_t \delta E \sim -i\Omega \delta E$), while the subsequent terms represent the diamagnetic and ponderomotive drifts.

Using Faraday's law~\eqref{eq:farday} and Amp\'ere's law~\eqref{eq:ampere}, we get the following expression:
\begin{equation}
\partial_{xx}(\delta E_{1y}) - \frac{1}{c^2} \partial_{tt} (\delta E_{1y}) + \partial_{zz} (\delta E_{1y}) = \frac{4\pi n_0 e}{c^2} \left[ \partial_t \delta v_{1iy} - \partial_t \delta v_{1ey} \right].
\label{eq:msw_1}
\end{equation}

Here, we neglect the electron contribution to the perpendicular ponderomotive force, $F_{ex}$, in comparison to the ion term $F_{ix}$. In the low-frequency limit, the perpendicular oscillatory velocity is dominated by the $\mathbf{E} \times \mathbf{B}$ drift, which is species-independent consequently, the ponderomotive force scales directly with mass, rendering the electron term negligible ($F_{ex} \ll F_{ix}$).

The direct substitutions of Equations~\eqref{eq:msw_ey} and \eqref{eq:msw_iy} into Equation~\eqref{eq:msw_1} (with $\partial_t \to -i\Omega$) transform into the following form:
\begin{equation}
\left( 1 + \frac{c_s^2}{v_A^2} \right) \partial_{xx} (\delta E_{1y}) - \left( \frac{1}{c^2} + \frac{1}{v_A^2} \right) \partial_{tt} (\delta E_{1y}) + \partial_{zz} (\delta E_{1y}) = \frac{4\pi n_0 e}{c^2} \left( \frac{i\Omega c F_{ix}}{e B_0} \right).
\label{eq:msw_2}
\end{equation}

Utilizing the continuity equation and the momentum equation, and neglecting $1/c^2$ in comparison to $1/v_A^2$ and considering the negligible density variation along the $z$-direction, we can drop $\partial_{zz}\delta E_y$ from Equation~\eqref{eq:msw_2}, which reduces to
\begin{equation}
\left( \partial_{tt} - v_A^2(1 + 2\beta) \partial_{xx} \right) \frac{\delta n_1}{n_0} = -\frac{4\pi n_0 v_A^2}{B_0^2} \partial_x \left( F_{ix} \right).
\label{eq:msw_final}
\end{equation}

We now evaluate the $x$-component of the ion ponderomotive force. In the low-frequency KAW limit ($\omega \ll \omega_{ci}$), the dominant contribution arises from the parallel electric field component, as this term scales as $\omega^{-2}$ while perpendicular Reynolds stress contributions scale as $\omega^{-4}$ or weaker. The time-averaged ponderomotive potential per ion is
\begin{equation}
U_{pond,i}^\parallel = \frac{e^2|E_z|^2}{4m_i\omega^2}.
\label{eq:U_pond}
\end{equation}
From the linearized KAW equations in Equation~\eqref{eq:Ez_from_By}, the parallel electric field scales as $E_z \sim (c\omega k_{0x}/\omega_{pe}^2)B_y$. Substituting into Equation~\eqref{eq:U_pond}, we find that the $\omega^2$ factors cancel, giving the ponderomotive force per unit volume:
\begin{equation}
F_{ix} = -n_0\partial_x (U_{pond,i}^\parallel) = -\frac{n_0 e^2c^2k_{0x}^2}{4m_i\omega_{pe}^4}\partial_x |B_y|^2.
\label{eq:Fix_pond}
\end{equation}

Substituting Equation~\eqref{eq:Fix_pond} into Equation~\eqref{eq:msw_final}, we obtain
\begin{equation}
\left( \partial_{tt} - v_A^2(1 + 2\beta) \partial_{xx} \right) \frac{\delta n_1}{n_0} = \frac{\pi n_0^2 v_A^2 e^2c^2k_{0x}^2}{m_i\omega_{pe}^4 B_0^2} \partial_{xx} |B_y|^2,
\label{eq:msw_intermediate}
\end{equation}

For density perturbations evolving on timescales much longer than the KAW period, the second time derivative becomes negligible compared to the spatial gradient terms. This adiabatic limit is appropriate when the density cavity adjusts quasi-statically to the slowly varying wave envelope. Neglecting $\partial_{tt}(\delta n_1/n_0)$ and integrating Equation~\eqref{eq:msw_intermediate} twice with respect to $x$, we find

\begin{equation}
\frac{\delta n_1}{n_0} = -\frac{m_e \lambda_e^2 k_{0x}^2}{16\pi m_i^2 v_A^2(1 + 2\beta)}|B_y|^2.
\label{eq:density_pert}
\end{equation}

The negative sign indicates that the ponderomotive force expels plasma from regions of large wave amplitude, creating a localized density depletion. Equation~\eqref{eq:density_pert} reveals several key scaling properties: the density perturbation scales as $(\lambda_ek_{0x})^2$, demonstrating that shorter electron inertial scales and larger parallel wavenumbers produce stronger ponderomotive coupling. The $(1+2\beta)^{-1}$ factor indicates that higher plasma beta weakens the density response through increased magnetosonic wave stiffness, as the enhanced thermal pressure provides greater resistance to plasma expulsion. This density cavity, in turn, modifies the local Alfv\'en speed and refractive index experienced by the KAW, establishing the nonlinear feedback necessary for wave packet self-interaction \citep{bellan1998fine,stasiewicz2000small}.

Substituting Equation \eqref{eq:density_pert} into the total density perturbation ($\delta n/n_0$), Equation \eqref{eq:density} can be now re-written as
\begin{equation}\label{eq:density_tot}
    \frac{\delta n}{n_0}= \frac{\delta n_1}{n_0} + \frac{\delta n_2}{n_0} = -\frac{m_e \lambda_e^2 k_{0x}^2}{16\pi m_i^2 v_A^2(1 + 2\beta)}|\tilde{B_y}|^2 + \eta\cos{(\alpha_z z)}.
\end{equation}

Upon substitution of Equation \eqref{eq:density_tot} into Equation \eqref{eq:kaw_envelope}, the following envelope equation is obtained for the evolution of KAW:
\begin{eqnarray}
&& 2\left(\gamma_L - i\omega - i\lambda_e^2\omega k_{0x}^2 + \gamma_L\lambda_e^2 k_{0x}^2\right) \partial_t \delta \tilde{B}_y
+ 2\left(i\omega^2\lambda_e^2 k_{0x} - 2\gamma_L\lambda_e^2\omega k_{0x} - ik_{0x} k_{0z}^2 v_{te}^2\lambda_e^2\right) \partial_x \delta\tilde{B}_y
\nonumber \\ &&
+ \left(\omega^2\lambda_e^2 + 2i\omega\gamma_L\lambda_e^2 - v_{te}^2 \lambda_e^2 k_{0z}^2\right) \partial_{xx} \delta\tilde{B}_y
+ 2\bigg[\frac{ik_{0z} B_0 c\beta \omega_{ci}}{8\pi n_0 e} 
- ik_{0x}^2 k_{0z} v_{te}^2\lambda_e^2 
\nonumber \\ &&
- k_{0z} v_A^2\Bigl(1 +\frac{m_e \lambda_e^2 k_{0x}^2}{16\pi m_i^2 v_A^2(1 + 2\beta)}|\tilde{B_y}|^2 - \eta\cos{(\alpha_z z)}\Bigr) \bigg]
\partial_z \delta \tilde{B}_y - \bigg[ k_{0x}^2 v_{te}^2\lambda_e^2 - \frac{B_0 c\beta \omega_{ci}}{8\pi n_0 e} 
\nonumber \\ &&
+ v_A^2 \Bigl(1 +\frac{m_e \lambda_e^2 k_{0x}^2}{16\pi m_i^2 v_A^2(1 + 2\beta)}|\tilde{B_y}|^2 - \eta\cos{(\alpha_z z)}\Bigl) \bigg]
\partial_{zz} \delta \tilde{B}_y - 4\left(\omega k_{0x} \lambda_e^2 + i\gamma_L\lambda_e^2 k_{0x}\right) \partial_{tx} \delta \tilde{B}_y
\nonumber \\ &&
+ 2\left(i\omega\lambda_e^2 - \gamma_L\lambda_e^2\right) \partial_{txx} \delta \tilde{B}_y
- 4 k_{0x} k_{0z} v_{te}^2\lambda_e^2 \partial_{xz} \delta \tilde{B}_y
+ 2i v_{te}^2 \lambda_e^2 k_{0z} \partial_{xxz} \delta \tilde{B}_y
+ v_{te}^2\lambda_e^2 \partial_{xxzz} \delta \tilde{B}_y
\nonumber \\ &&
+ \bigg[ v_A^2k_{0z}^2\Bigl(1 + \frac{m_e \lambda_e^2 k_{0x}^2}{16\pi m_i^2 v_A^2(1 + 2\beta)}|\tilde{B_y}|^2 - \eta\cos{(\alpha_z z)}\Bigl)
- \frac{B_0 c\beta \omega_{ci} k_{0z}^2}{8\pi n_0 e} + k_{0x}^2 k_{0z}^2 v_{te}^2\lambda_e^2
\nonumber \\ &&
- 2i\omega\gamma_L\lambda_e^2 k_{0x}^2 - \lambda_e^2\omega^2 k_{0x}^2 - \omega^2 - 2i\omega\gamma_L \bigg] \delta \tilde{B}_y = 0
\label{eq:kaw_envelope2}
\end{eqnarray}

Normalizing Equation \eqref{eq:kaw_envelope2}, we obtain the following dimensionless equation,
\begin{eqnarray}
&&
c_1\!\left(\tfrac{\gamma_L}{\omega}-i\right)\partial_t \delta B_y
+ \left(i c_2-\tfrac{2\gamma_L}{\omega}\right)\partial_x \delta B_y
+ c_3\!\left(1+2i\tfrac{\gamma_L}{\omega}-\tfrac{v_{te}^2}{v_A^2}\right)\partial_{xx}\delta B_y
+ i c_4\!\left\{1+|\delta B_y|^2-\eta\cos(\alpha_z z)\right\}\partial_z \delta B_y
\nonumber \\[1ex]
&&
- c_5\!\left\{1+|\delta B_y|^2-\eta\cos(\alpha_z z)\right\}\partial_{zz}\delta B_y
- c_6\!\left(1+i\tfrac{\gamma_L}{\omega}\right)\partial_{tx}\delta B_y
+ c_7\!\left(i-\tfrac{\gamma_L}{\omega}\right)\partial_{txx}\delta B_y
- c_8\partial_{xz}\delta B_y
+ i c_9 \partial_{xxz}\delta B_y
\nonumber \\[1ex]
&&
+ c_{10}\partial_{xxzz}\delta B_y
+ \left[\!\left\{1+|\delta B_y|^2-\eta\cos(\alpha_z z)\right\}
- c_{11} - i c_{12}\tfrac{2\gamma_L}{\omega}\right]\delta B_y = 0 .
\label{eq:final_kaw}
\end{eqnarray}

Here, the dimensionless coefficients $c_1, c_2, ..., c_{12}$ are defined as follows:
\begin{eqnarray*}
&&c_1 = 1 + \lambda_e^2 k_{0x}^2, ~~
c_2 = 1 - \frac{k_{0z}^2 v_{te}^2} {\omega^2}, ~~
c_3 = \frac{v_A^2k_{0z}^2}{4\omega^2\lambda_e^2 k_{0x}^2}, ~~
c_4 = \left( \frac{\beta v_A^2}{2k_{0x}^2 v_{te}^2\lambda_e^2} - \frac{v_A^2}{k_{0x}^2v_{te}^2\lambda_e^2} \frac{\omega_{ci}^2-\omega^2}{\omega_{ci}^2} - 1 \right), \\
&&c_5 = \frac{v_A^2}{4k_{0x}^4v_{te}^4\lambda_e^4}\left[k_{0x}^2v_{te}^2\lambda_e^2 - \frac{\beta v_A^2}{2} + v_A^2\left(\frac{\omega_{ci}^2-\omega^2}{\omega_{ci}^2}\right)\right], ~~
c_6 = \frac{v_A^2k_{0z}^2}{\omega^2}, ~~
c_7 = \frac{v_A^4k_{0z}^4}{4\omega^4\lambda_e^2k_{0x}^2}, ~~
c_8 = \frac{v_A^2k_{0z}^2}{\omega^2\lambda_e^2k_{0x}^2}, \\
&&c_9 = \frac{v_A^4k_{0z}^4}{8\omega^4\lambda_e^4k_{0x}^4}, ~
c_{10} = \frac{(1 + 2\beta)(1 + \lambda_e^2 k_{0x}^2) v_A^6 k_{0z}^4}{16 \omega^4 \lambda_e^6 k_{0x}^6 v_{te}^2}, ~
c_{11} = \frac{\beta}{2} - \frac{k_{0x}^2v_{te}^2\lambda_e^2}{v_A^2} + \frac{\lambda_e^2\omega^2k_{0x}^2}{v_A^2k_{0z}^2} + \frac{\omega^2}{v_A^2k_{0z}^2},~
c_{12} = \frac{\omega^2 \lambda_e^2k_{0x}^2 + \omega^2}{v_A^2k_{0z}^2}
\end{eqnarray*}

The normalization parameters are
$t_N = \frac{2\omega}{v_A^2k_{0z}^2}$,
$x_N = \frac{2\omega^2 \lambda_e^2k_{0x}}{v_A^2k_{0z}^2}$,
$z_N = \frac{2k_{0x}^2v_{te}^2\lambda_e^2}{v_A^2k_{0z}}$, and
$B_N=\frac{4m_iv_A}{\lambda_ek_{0x}}\sqrt{\frac{\pi(1 + 2\beta)}{m_e}}$

The Landau damping rate $\gamma_L$ for KAWs in finite-$\beta$ plasmas is calculated using \citet{lysak1996kinetic} and \citet{hasegawa1976particle}:
\begin{equation}
\label{eq:gamma_L}
\frac{\gamma_L}{\omega} \approx \sqrt{\pi} \frac{v_A k_x^2 \rho_i^2}{v_{te}},
\end{equation}
where $k_x$ is the perpendicular wavenumber generated by nonlinear interaction of pump KAW and MSW. This expression captures the scale-dependent nature of Landau damping, with dissipation increasing as $k_x^2$ at smaller scales.

\section{Numerical Methodology}
\label{sec:numerical_simulation}

By treating Equation \eqref{eq:final_kaw} as a modified nonlinear Schr\"odinger equation that describes the dispersive and nonlinear properties of KAWs, we employ a pseudospectral method based on Fast Fourier Transforms (FFTs) for spatial discretization. In this approach, spatial derivatives $\partial/\partial x$ and $\partial/\partial z$ are converted to the wavenumber domain by multiplying with $ik_x$ and $ik_z$, where $k_x$ and $k_z$ are the wavenumbers corresponding to the $x$ and $z$ directions, respectively. The nonlinear terms such as $|\delta B_y|^2$ and their coupling with spatial derivatives are first evaluated in real space, then transformed back to Fourier space using FFTs. This hybrid approach avoids expensive convolution operations that would be required if nonlinear products were computed directly in Fourier space. The transformation of spatial derivatives to the spectral domain effectively reduces the partial differential equation (PDE) to a system of ordinary differential equations (ODEs) in time, which can then be solved using standard time integration methods.

We solve Equation \eqref{eq:final_kaw} numerically by Adams--Bashforth predictor and Adams--Moulton correcter method. Since this is a multi step method, initially, we find the numerical solution for the first four time steps by the fourth-order Runge--Kutta (RK4) method. To address aliasing errors from nonlinear terms in our pseudospectral method, we employ a standard zero-padding technique. The spectral grid is extended from $256$ to $512$ modes in each direction by padding Fourier coefficients with zeros. Nonlinear products are then computed on the enlarged physical grid, transformed back to Fourier space, and truncated to the original $256$ modes. This padding 
factor of 2 exceeds the theoretical minimum of 3/2 for quadratic nonlinearities \citep{liu2012controlling}, ensuring complete de-aliasing at the cost of increased computational expense. The implementation of our time integration scheme is complicated by mixed derivative terms in the dynamical equation, such as $\partial_t\partial_x \delta B_y$ and $\partial_t\partial_{xx} \delta B_y$. To handle these terms, we rearrange the equation to isolate the pure time derivative
\( c_1\Bigl(\frac{\gamma_L}{\omega}-i\Bigr)\,\partial_t \delta B_y \)
on the left-hand side, while moving all other terms (spatial derivatives, nonlinear interactions, and mixed derivatives) to the right-hand side. Since these terms couple spatial and temporal evolution, we evaluate the time derivatives contained within them numerically. Specifically, we employ a finite difference approximation using field values from the current and provisional (predicted) time steps. This procedure converts the implicit mixed terms into explicit spatial functions, allowing the system to be advanced via the predictor-corrector scheme.

Simulations are carried out on a uniform $256\times 256$ grid with periodic boundary conditions in both directions. The computational domain is square, with $L_x = 2\pi/\alpha_x$ and $L_z = 2\pi/\alpha_z$, where the characteristic perturbation scale is set to $\alpha_x = \alpha_z = 0.5$. Time integration uses a fixed timestep $\Delta t = 1\times10^{-5}$ (in normalized unit).

To implement these algorithms efficiently, the numerical implementation is written in \texttt{Fortran 90}, utilizing modules and subroutines from the \textit{Numerical Recipes in Fortran 90} library \citep{press1996numerical}, including \texttt{nrtype.f90}, \texttt{nrutil.f90}, \texttt{nr.f90}, and the FFT routine \texttt{four2(data,isign)}. To validate our numerical approach, we first developed and tested an algorithm for the two-dimensional cubic NLSE by comparing results with previous studies. We verified conservation of the plasmon number to within $10^{-5}$ accuracy, defined as
\begin{equation}
P=\frac{1}{L_xL_z}\int_0^{L_x}\int_0^{L_z}|\delta B_{y}|^2\,dx\,dz
=\int_{-\infty}^{\infty}\int_{-\infty}^{\infty} |\delta B_{yk}|^2\,dk_{x}\, dk_{z}
=\sum_{k}|\delta B_{yk}|^2.
\end{equation}
This validated algorithm was then adapted to solve the modified NLSE-type system governing our non-integrable problem. The initial condition used in our simulation consists of a uniform plane pump KAW of fixed amplitude, superimposed by sinusoidal perturbations in both spatial directions, given by
\begin{equation}
 \delta B_y(x,z,t=0) = \delta B_{y0}\left[1+\epsilon \cos(\alpha_x x)\right] \left[1+\epsilon \cos(\alpha_z z)\right],
\end{equation}
where $\delta B_{y0}=1$ is the pump amplitude, $\epsilon =0.1$ is the perturbation magnitude, and $\alpha_x=\alpha_z=0.5$. While magnetic fluctuations in space plasmas are complex, this initial condition remains physically relevant as a superposition of fundamental wave components.

We adopt plasma parameters representative of Earth's magnetosheath from \citet{artemyev2022ion}: $B_0\approx1.50\times10^{-4}$~G, $n_0\approx9.00$~cm$^{-3}$, $T_e\approx1.47 \times 10^5$~K, and $T_i\approx1.62 \times 10^6$~K. Derived parameters include plasma beta $\beta \approx 2.45$, frequencies $\omega_{ce}\approx8.79\times10^2$~rad~s$^{-1}$, $\omega_{ci}\approx0.48$~rad~s$^{-1}$, and $\omega_{pe}\approx1.69\times10^5$~rad~s$^{-1}$, and the electron skin depth $\lambda_e\approx1.77 \times 10^5$~cm. The simulation employs a frequency ratio $\omega/\omega_{ci}=0.1$ and normalized wavenumber $k_{0x}\lambda_e =0.2$. All normalizations and coefficients appearing in Equation \eqref{eq:final_kaw} are computed from these physical parameters and summarized in Table~\ref{tab1}.

\begin{table}[h!]
\centering
\caption{Simulation parameters and dimensionless coefficients for typical magnetosheath plasma.}
\label{tab1}
\begin{tabular}{l c | l c}
\hline
Parameter & Value & Coefficient & Value \\
\hline
$\rho_i$ & $1.1382\times10^7~\mathrm{cm}$ & $c_1$ & $1.0400$ \\
$\rho_s$ & $1.4691\times10^8~\mathrm{cm}$ & $c_2$ & $-1.3278$ \\
$v_A$ & $1.0906\times10^{7}~\mathrm{cm/s}$ & $c_3$ & $0.0388$ \\
$c_s$ & $1.1564\times10^{7}~\mathrm{cm/s}$ & $c_4$ & $-1$ \\
$v_{te}$ & $2.1109\times10^{8}~\mathrm{cm/s}$ & $c_5$ & $0.0164$ \\
$k_{0z}$ & $1.0384\times10^{-9}~\mathrm{cm^{-1}}$ & $c_6$ & $0.00621$ \\
$k_{0x}$ & $1.1290\times10^{-6}~\mathrm{cm^{-1}}$ & $c_7$ & $0.00024$ \\
$x_N$ & $1.1403\times10^{7}~\mathrm{cm}$ & $c_8$ & $0.1553$ \\
$z_N$ & $1.4430\times10^{10}~\mathrm{cm}$ & $c_9$ & $0.0030$ \\
$t_N$ & $31.60~\mathrm{s}$ & $c_{10}$ & $0.00062$ \\
$B_N$ & $5.2\times10^{-5}~\mathrm{G}$ & $c_{11}$ & $153.61$ \\
$\delta B_{y0}$ & $1.0$ & $c_{12}$ & $167.37$ \\
$\epsilon$ & $0.1$ & $\alpha_x$ & $0.5$ \\
$\eta$ & $0.1$ & $\alpha_z$ & $0.5$ \\
 & & $\Delta t$ & $1\times10^{-5}$ \\
\hline
\end{tabular}
\end{table}

\section{Results and Discussion} \label{sec:results}

\begin{figure}[ht!]
    \centering
    \includegraphics[width=0.6\linewidth]{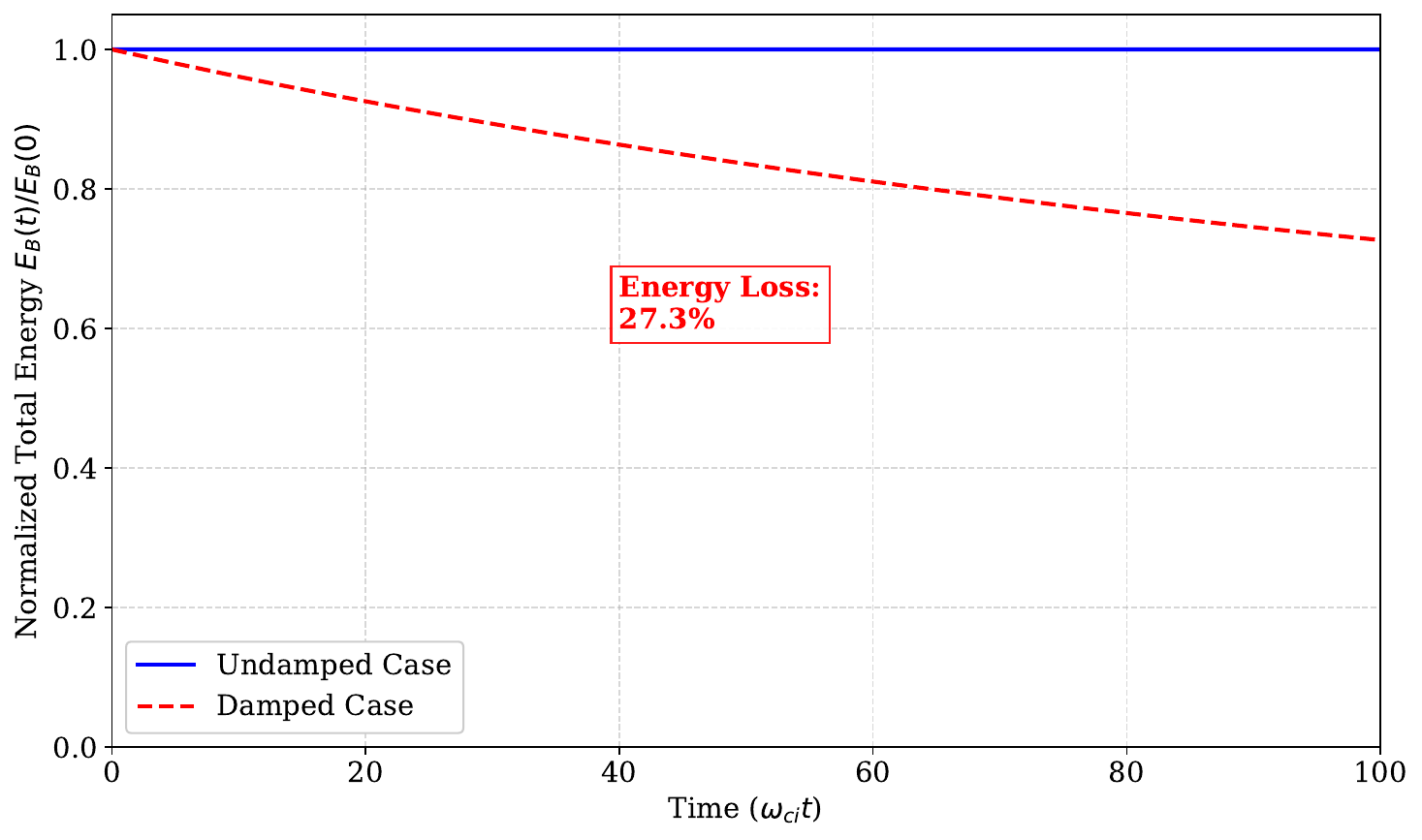}
    \caption{Time evolution of normalized total magnetic energy $E_B(t)/E_B(0)$. The undamped case (blue solid line) conserves energy, while the Landau-damped case (red dashed line) shows monotonic decay with $27.3\%$ energy loss over $100\,\omega_{ci}^{-1}$.}
    \label{fig:energy_decay}
\end{figure}

To validate the numerical stability of our code and quantify energy dissipation due to Landau damping, we examine the time evolution of the total magnetic energy. Figure~\ref{fig:energy_decay} shows the normalized total magnetic energy, $E_B(t)/E_B(0)$, integrated over the simulation domain for both cases. The undamped case (blue solid line) maintains constant energy throughout the simulation time ($\omega_{ci}t = 0$ to $100$). This confirms that our pseudospectral scheme is numerically stable and free from artificial dissipation. The energy conservation provides a reliable baseline, ensuring that any dynamics observed are physically driven rather than numerical artifacts. In contrast, the Landau-damped case (red dashed line) exhibits monotonic energy decay. By $\omega_{ci}t = 100$, the system has dissipated approximately $27.3\%$ of its initial magnetic energy. This steady energy loss demonstrates that the damping term effectively represents the physical process of wave energy transfer to plasma particles via wave-particle interactions.

\begin{figure}[ht!]
    \centering
    \includegraphics[width=1.0\linewidth]{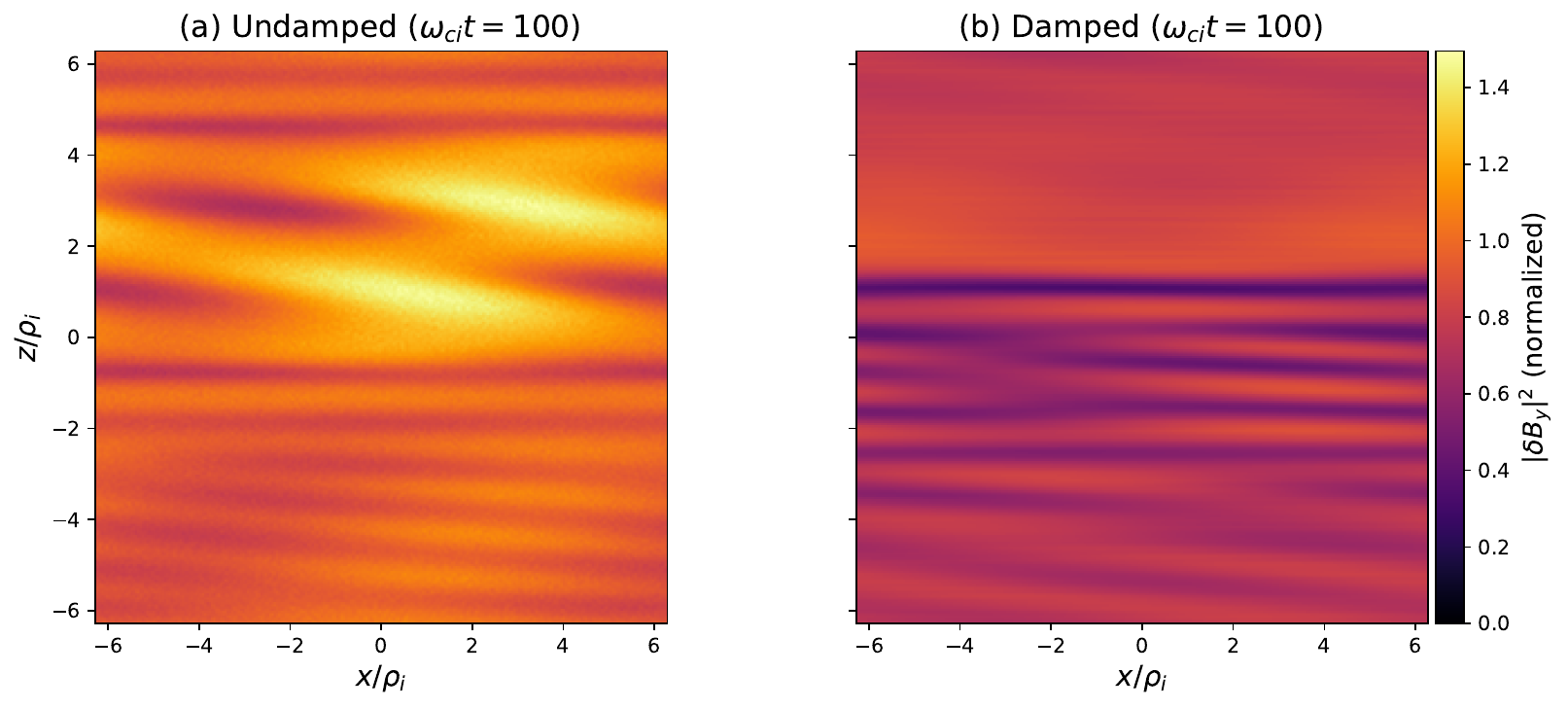}
    \caption{Spatial distribution of normalized magnetic field intensity $|\delta B_y|^2$ at $\omega_{ci}t = 100$. (a) Undamped case showing intense filamentary structures. (b) Damped case showing suppressed small-scale features.}
    \label{fig:morphology}
\end{figure}

Figure~\ref{fig:morphology} shows the spatial distribution of magnetic field intensity $|\delta B_y|^2$ at $\omega_{ci}t = 100$, illustrating the structural differences caused by Landau damping. In the undamped case (panel a), the field exhibits intense, filamentary structures oriented perpendicular to the background field. The field intensity varies by more than an order of magnitude across the domain, with peak values reaching approximately 1.4 times the initial amplitude. These sharp gradients indicate strong nonlinear wave steepening. The damped case (panel b) presents a noticeably different structure. The filamentary structures are largely suppressed, and the field displays a smoother, more uniform distribution. Peak intensities are reduced to approximately 1.0, and the spatial variations are less pronounced. This smoothing results from Landau damping acting preferentially on shorter wavelength perturbations. The contrast between panels (a) and (b) demonstrates that collisionless damping fundamentally affects the spatial structure of the turbulence, not just the overall energy level. This smoothing of spatial structures is consistent with the energy dissipation shown in Figure~\ref{fig:energy_decay}.

\begin{figure}[ht!]
    \centering
    \includegraphics[width=1.0\linewidth]{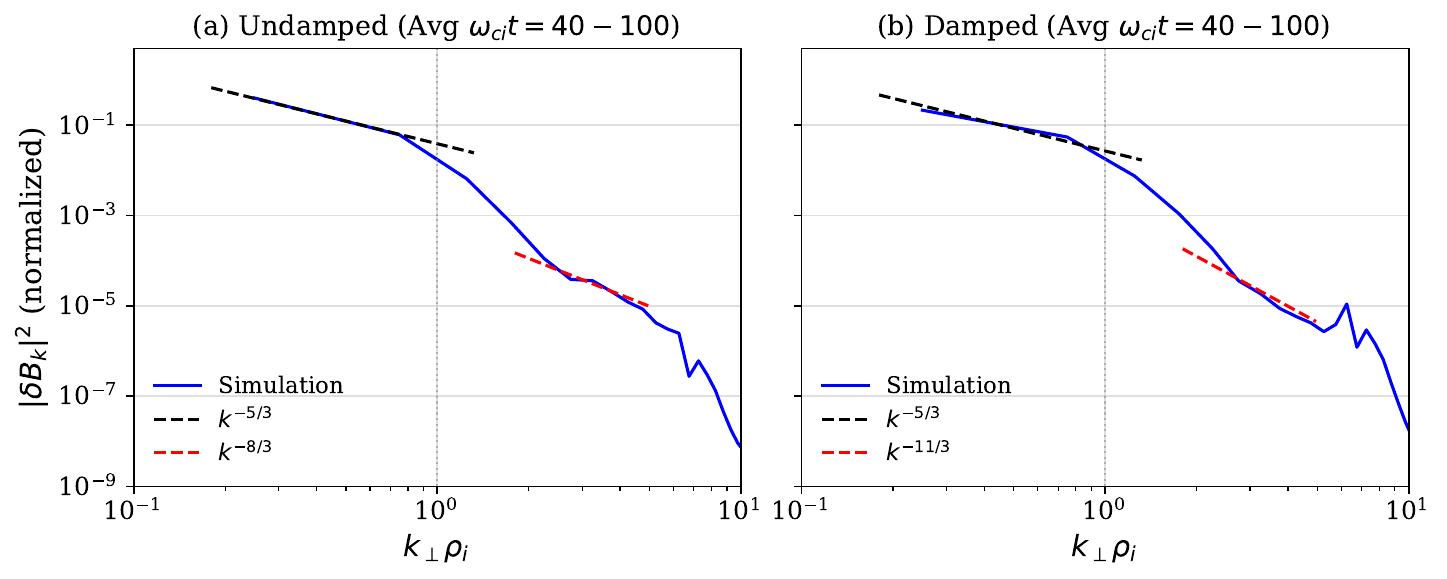}
    \caption{Time-averaged magnetic power spectra for $\omega_{ci}t = 40$--$100$. (a) Undamped case showing a $k^{-8/3}$ kinetic range. (b) Damped case showing a steeper $k^{-11/3}$ dissipation range due to Landau damping.}
    \label{fig:spectra}
\end{figure}

Figure~\ref{fig:spectra} presents the time-averaged magnetic power spectra for $\omega_{ci}t = 40$--$100$, illustrating how Landau damping modifies the turbulent cascade. In the undamped regime (panel a), the spectrum exhibits a $k_{\perp}^{-5/3}$ scaling in the inertial range ($k_\perp\rho_i < 1$), consistent with the Goldreich-Sridhar model for anisotropic MHD turbulence \citep{goldreich1995toward}. At sub-ion scales ($k_\perp\rho_i > 1$), the spectrum transitions to a steeper scaling of approximately $k_{\perp}^{-8/3}$ ($\approx -2.67$). This slope characterizes the dispersive KAW cascade in the absence of significant collisionless dissipation \citep{boldyrev2012spectrum, sahraoui2009evidence}. In the damped regime (panel b), the large-scale inertial range retains the $k_\perp^{-5/3}$ scaling, indicating that kinetic-scale dissipation does not significantly affect the energy injection scales. However, for $k_\perp\rho_i > 1$, the spectrum steepens sharply to $k_{\perp}^{-11/3}$ ($\approx -3.67$). This steeper decay reflects the efficiency of electron Landau damping in removing energy from the cascade before it reaches electron scales \citep{alexandrova2012solar, passot1993multidimensional}. Together, these two cases establish theoretical bounds for the kinetic spectral slope: a shallow limit of $-8/3$ for weak damping and a steep limit of $-11/3$ for a significantly high damping.

\begin{figure}[ht!]
    \centering
    \includegraphics[width=1.0\linewidth]{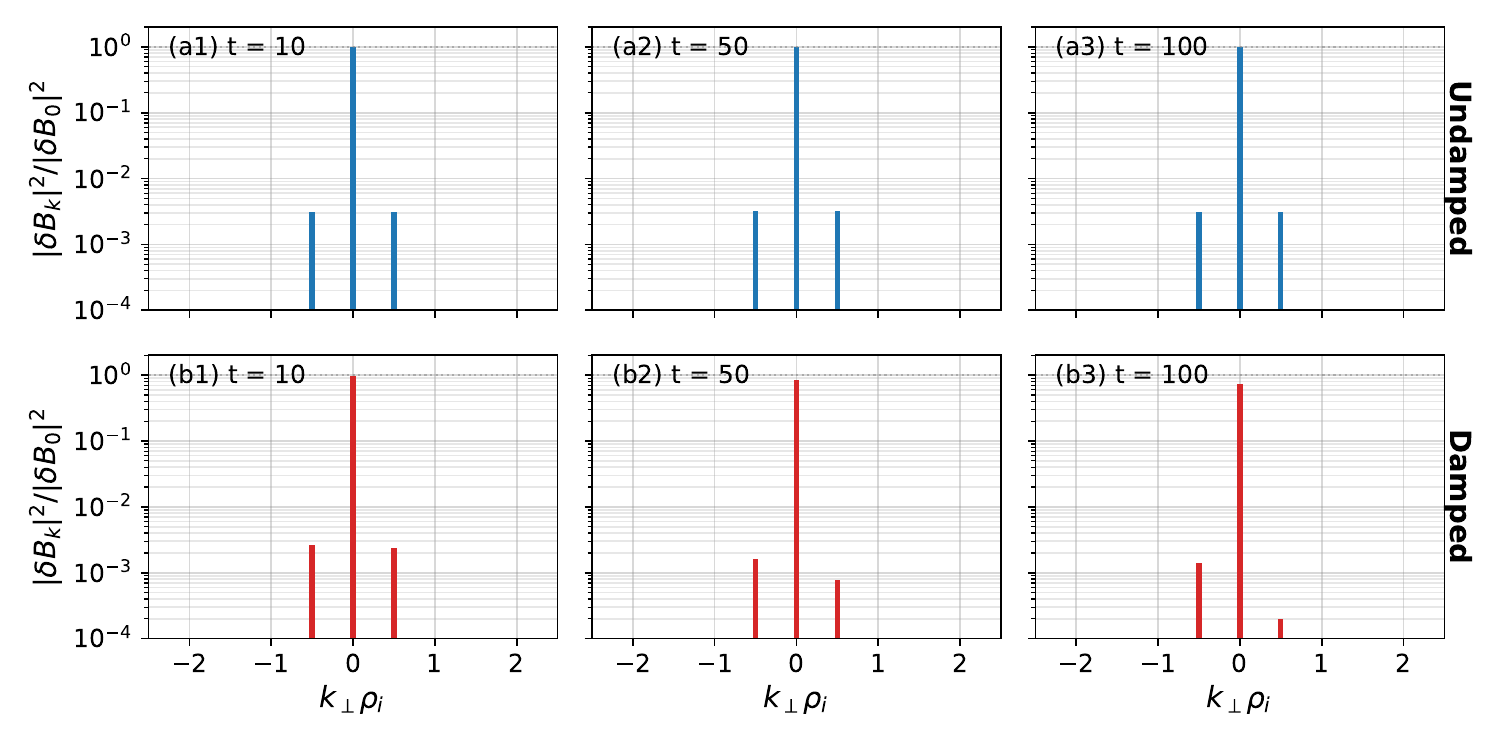}
    \caption{Time evolution of discrete Fourier mode amplitudes normalized by initial pump energy ($|\delta B_k|^2/|\delta B_0|^2$). Top row (a1--a3): Undamped case showing persistent pump mode. Bottom row (b1--b3): Damped case showing pump mode decay due to Landau damping.}
    \label{fig:sticks}
\end{figure}

Figure~\ref{fig:sticks} shows the evolution of discrete Fourier modes to illustrate how damping affects different wavenumbers. The spectral power is normalized by the initial pump amplitude $|\delta B_0|^2$, allowing direct comparison of energy redistribution between the undamped and damped cases. In the undamped simulation (top row), the fundamental pump mode at $k_\perp\rho_i = 0$ remains at unity throughout the simulation. Energy transfers to the sidebands at $k_\perp\rho_i \approx \pm 0.5$--$0.6$ through modulational instability, but the pump mode itself experiences no depletion. This is consistent with purely conservative nonlinear dynamics, where energy redistributes among modes without net loss. In contrast, the damped simulation (bottom row) shows different behavior. Initially, the fundamental mode at $k_\perp\rho_i=0$ remains at unity (panel b1), but at $\omega_{ci}t = 100$ (panel b3), the pump mode amplitude has decreased to approximately 0.7, indicating direct energy removal from the fundamental scale. This demonstrates that Landau damping does not only act on high-wavenumber fluctuations; it also extracts energy directly from the pump mode at the energy-containing scales. The sideband amplitudes in panel (b3) are also suppressed compared to panel (a3), confirming that damping regulates energy across all active modes. This multi-scale energy removal distinguishes collisionless damping from purely cascade-driven dissipation, where energy would only be removed at the smallest scales.

\section{Comparison with MMS Spacecraft Observations}\label{sec:comparision}

The MMS mission, launched by NASA in March 2015, has been important for understanding plasma dynamics throughout Earth's magnetospheric system, including the magnetosheath. The mission comprises four identical spacecraft that fly in a tetrahedral formation, allowing simultaneous multipoint measurements of electromagnetic fields and plasma parameters at spatial separations down to electron kinetic scales. In burst mode, the FluxGate Magnetometer (FGM) samples the vector magnetic field at 128 Hz, the Electric Double Probe (EDP) measures the full three-component electric field at comparable rates, and the Fast Plasma Investigation (FPI) provides ion and electron moments with sampling rates up to 30 ms \citep{burch2016magnetospheric,russell2016magnetospheric,pollock2016fast,allmann2021fluid,williams2025magnetospheric}.

To validate our numerical simulation results of KAW dynamics with in situ MMS observations, we select MMS1 Level-2 data from 28 December 2015 01:48:00--01:53:00 UT, when the spacecraft traversed the dayside magnetosheath at a plasma $\beta \approx 2$--$3$ \citep{stawarz2023preface}. This interval has been well studied in prior MMS analyses, exhibiting pronounced Alfv\'enic fluctuations and steepened magnetic spectra near ion scales \citep{macek2018magnetospheric,macek2023statistical}. For observational data analysis and visualization, we utilize the Python-based Space Physics Environment Data Analysis Software (PySPEDAS) library, which provides standardized access to various spacecraft data including MMS data and comprehensive analysis tools \citep{angelopoulos2019space,grimes2022space}. Through PySPEDAS, we retrieve the FGM, EDP, and FPI data exclusively in burst mode to ensure high temporal resolution and precise synchronization. We utilize the FGM's native 128 Hz sampling for magnetic spectral analysis and concurrent high-cadence FPI ion moments to characterize the background plasma conditions (e.g., density and flow velocity) necessary for accurate Alfv\'en speed calculations. The retrieved data are processed using standard PySPEDAS quality control procedures, including automated spike removal and coordinate transformations from the spacecraft's native Despun Sun-Locked (DSL) system to the Geocentric Solar Ecliptic (GSE) coordinate system \citep{torbert2016fields,grimes2022space}. Plasma conditions during the selected interval are verified to confirm consistency with our simulation parameter space. The power spectral densities of the burst-mode magnetic field are computed using Welch's method \citep{welch1967use} with Hanning windowing and linear detrending to reduce spectral leakage and provide accurate frequency-domain characterization of the magnetic field fluctuations \citep{harris1978use}.


\begin{figure*}[ht!]
  \centering
  \includegraphics[width=0.9\textwidth]{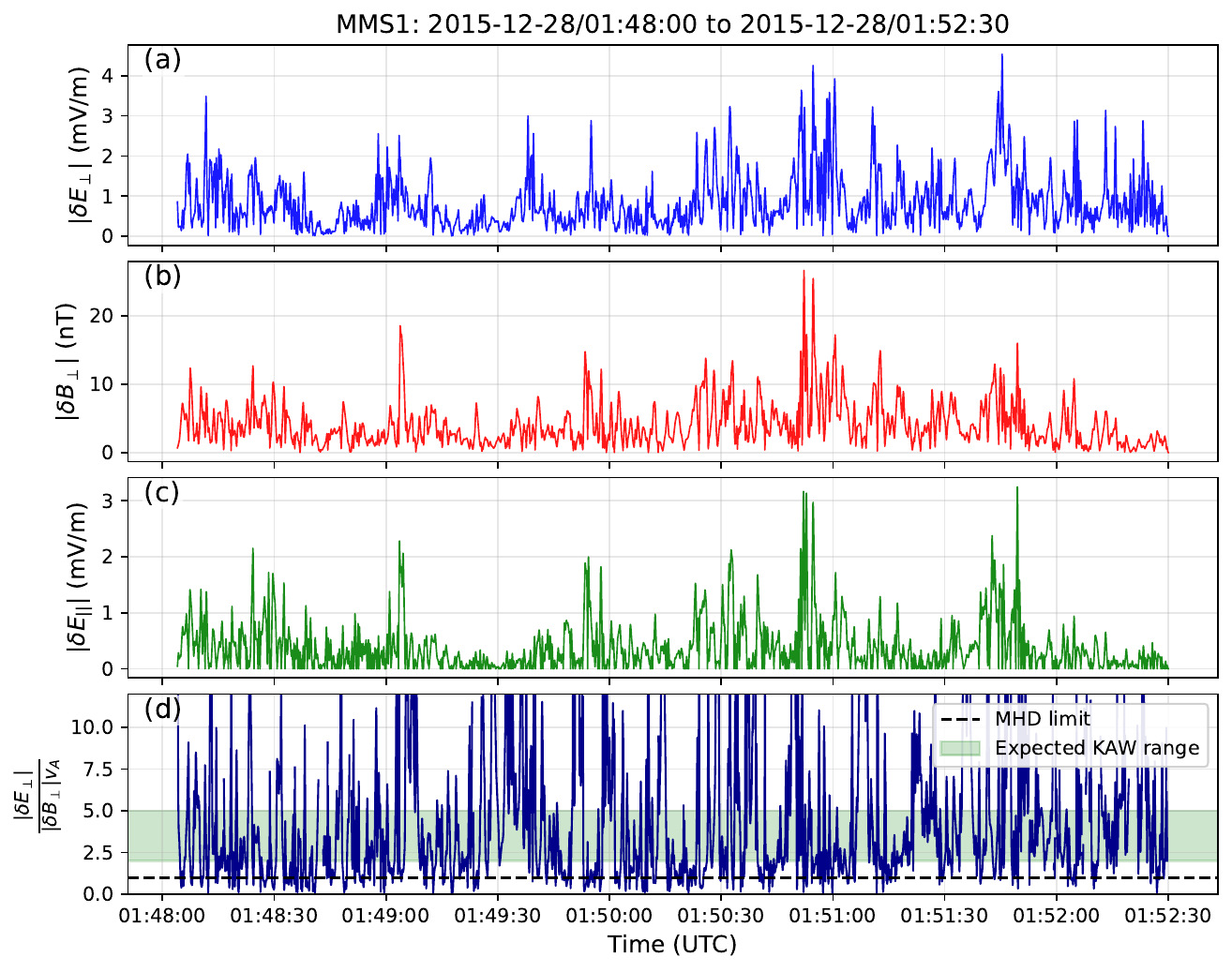}
  \caption{MMS1 electromagnetic field analysis. Top three panels show the amplitudes of (a) perpendicular electric field $|\delta E_\perp|$, (b) perpendicular magnetic field $|\delta B_\perp|$, and (c) parallel electric field $|\delta E_\parallel|$. (d) The normalized ratio $|\delta E_\perp|/(|\delta B_\perp|v_A)$. The black dashed line marks the ideal MHD limit (ratio $\approx 1$), while the green shaded region indicates the expected range for KAWs.}
   \label{fig:mms_kaw_comprehensive}
\end{figure*}

Figure~\ref{fig:mms_kaw_comprehensive} presents electromagnetic field measurements from MMS1 obtained during a 4.5-minute interval in the dayside magnetosheath on December 28, 2015. The perpendicular electric field fluctuations (Panel a) range from 0.2--4~mV/m, while magnetic fluctuations (Panel b) span 2--25~nT, showing correlated variations across the measurement period. These observations reveal characteristics consistent with KAWs and provide clear evidence for deviations from ideal MHD behavior \citep{hollweg1999kinetic,stasiewicz2000small}. To quantify these deviations, we calculated the normalized ratio of perpendicular electric to magnetic field fluctuations, $|\delta E_\perp|/(|\delta B_\perp|v_A)$, shown in Panel (d). This ratio consistently exceeds the ideal MHD limit of 1 (indicated by the black dashed line). The values frequently fall within or exceed the expected range for KAWs (green shaded region), serving as a key observational signature for distinguishing KAWs from MHD Alfv\'en waves in turbulent plasmas \citep{stasiewicz2000small,salem2012identification}. An important distinguishing feature is the presence of finite parallel electric field components ($|\delta E_\parallel| = 0.5$--2~mV/m) shown in Panel (c). This characteristic clearly distinguishes the observed fluctuations from ideal MHD Alfv\'en waves, which strictly require $\delta E_\parallel = 0$ \citep{lysak1996kinetic,wygant2002evidence}. The enhanced ratios and parallel fields occur intermittently in distinct wave packets, with particularly prominent examples around 01:50:00 and 01:51:00~UTC. This intermittency is characteristic of localized kinetic processes operating within magnetosheath turbulence \citep{roberts2018ion}. These measurements demonstrate that electromagnetic fluctuations in this region exhibit significant kinetic modifications at ion scales \citep{chen2013nature,podesta2012scale}.

\begin{figure}[ht!]
  \centering
  \includegraphics[width=0.8\textwidth]{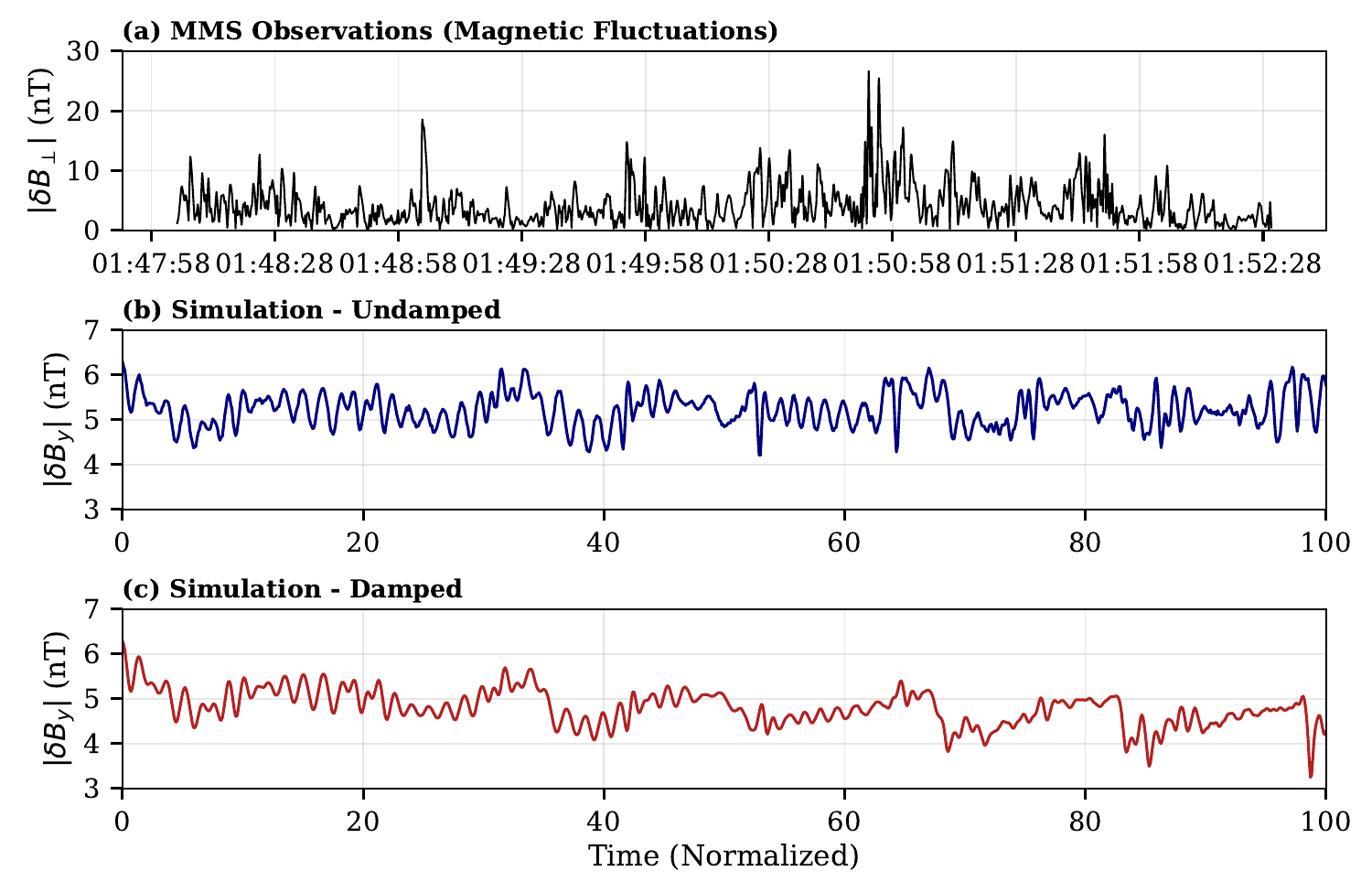}
  \caption{Comparison of magnetic field fluctuations. (a) MMS observations showing intermittent turbulent fluctuations ($|\delta B_{\perp}|$). (b) Undamped simulation showing sustained oscillations. (c) Landau-damped simulation showing amplitude suppression over time.}
  \label{fig:combined_magnetic}
\end{figure}

Figure~\ref{fig:combined_magnetic} compares magnetic field fluctuations observed by MMS with the temporal evolution of wave envelopes from our simulations. Panel (a) shows the MMS observations, which display highly intermittent behavior with fluctuation amplitudes reaching up to 30~nT. These bursty structures are characteristic of turbulence in the magnetosheath. Panel (b) presents the undamped simulation, which exhibits sustained oscillations with amplitudes around 4--6~nT. These quasi-periodic structures arise from modulational instability, where the initial wave packet breaks into localized magnetic envelopes. The coherent, wave-like character of the simulation differs from the stochastic nature of the MMS data, but both show the presence of large-amplitude magnetic structures. Panel (c) shows the damped case, where Landau damping suppresses the wave amplitude over time. By $\omega_{ci}t = 100$, the oscillations have weakened significantly compared to the undamped case. This demonstrates that collisionless damping regulates the wave energy and prevents sustained growth. It is important to note the different observational frames: the simulation tracks temporal evolution in the plasma rest frame over $\sim 100\,\omega_{ci}^{-1}$ (corresponding to several ion gyroperiods), while MMS captures a spatial snapshot as it moves through the turbulent region in $\sim 4$~minutes. Despite this difference, the simulation amplitudes are consistent with typical background fluctuation levels observed by MMS, supporting the role of Landau damping in limiting wave growth at kinetic scales.

\begin{figure}[ht!]
  \centering
  \includegraphics[width=0.8\textwidth]{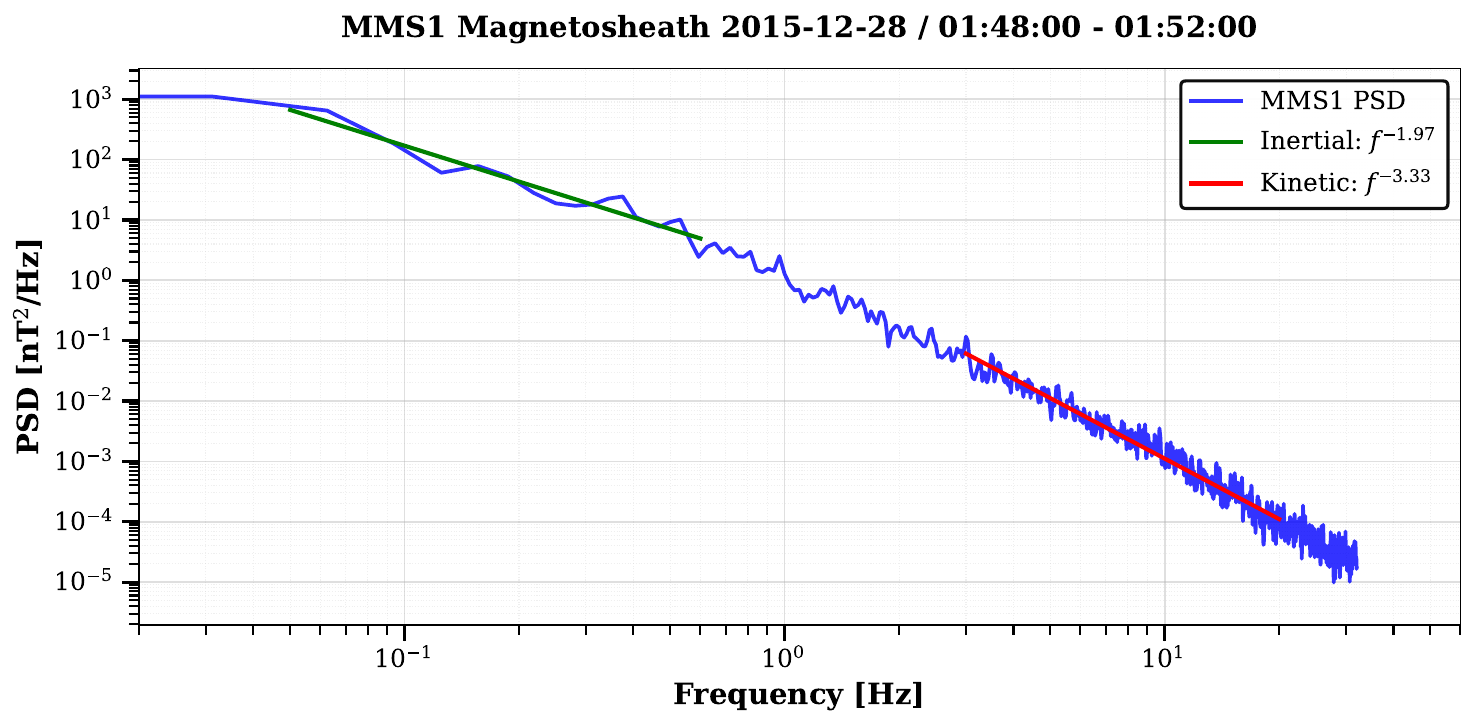}
  \caption{Power spectral density of magnetic field fluctuations from MMS1. The inertial range exhibits a steepened slope of $f^{-1.97}$ due to compressibility. The kinetic range slope of $f^{-3.33}$ lies within the theoretical bracket predicted by the simulations ($k^{-8/3}$ to $k^{-11/3}$), indicating active kinetic damping.}
  \label{fig:psd}
\end{figure}

Figure~\ref{fig:psd} displays the power spectral density of magnetic field fluctuations from MMS1 burst-mode data. The spectrum reveals a distinct spectral break near $f \approx 0.6$~Hz, separating the fluid and kinetic regimes. In the inertial range ($0.05 < f < 0.6$~Hz), we observe a spectral index of $f^{-1.97}$ (green line). This is notably steeper than the standard Kolmogorov $-5/3$ prediction recovered in our simulations. Such steepening is a well-documented feature of magnetosheath turbulence, where high compressibility, shocklet structures, and strong intermittency modify the energy transfer rate \citep{huang2017existence, hadid2017energy}.

To quantitatively compare the MMS frequency-domain observations with our wavenumber-domain simulations, we apply Taylor's frozen-in flow hypothesis \citep{taylor1938spectrum}. This hypothesis relates the spacecraft-frame frequency $f$ to the plasma-frame wavenumber through $k_\perp = 2\pi f/V_{flow}$, where $V_{flow}$ is the bulk plasma flow velocity. A key property of this transformation is that spectral indices are preserved: if $P(k) \propto k^\alpha$, then $P(f) \propto f^\alpha$. For Taylor's hypothesis to be valid, the bulk flow must significantly exceed the wave phase velocity, $V_{flow} \gg v_{phase}$. In the magnetosheath, typical flow velocities of 200--400~km/s yield $V_{flow}/v_A \approx 2$--$4$, satisfying this condition reasonably well \citep{huang2017existence,chen2013nature}. We first verify consistency of the spectral break location. The observed break at $f_{break} \approx 0.6$~Hz should correspond to the ion kinetic scale $k_\perp\rho_i \approx 1$. Using our simulation parameters ($\rho_i = 113.82$~km), we obtain:
\begin{equation}
V_{flow} = 2\pi f_{break} \rho_i = 2\pi \times 0.6 \times 113.82 \approx 430~\text{km/s}.
\label{eq:taylor}
\end{equation}
This value lies at the upper range of typical magnetosheath flows but is physically reasonable for the dayside region during the observed interval \citep{lucek2005magnetosheath, ma2020statistical, artemyev2022ion}.

Since spectral indices are preserved under Taylor's transformation, we can directly compare the kinetic range slopes. In the sub-ion range ($f > 3.0$~Hz), the MMS spectrum follows a power law of $f^{-3.33}$ (red line). This observed index falls between our undamped prediction ($-2.67$) and damped prediction ($-3.67$), indicating that the magnetosheath turbulence operates in an intermediate damping regime. The bracketing of the observed slope by the spectral indices from our limiting simulation cases provides strong evidence that Landau damping actively regulates the turbulent cascade at sub-ion scales.

\section{Conclusion} \label{conclusion}

This study shows that Landau damping controls the spectral properties and spatial structure of KAW turbulence in Earth's magnetosheath. By solving a modified nonlinear Schr\"odinger equation and comparing the results with MMS spacecraft observations, we have shown how collisionless dissipation modifies the turbulent cascade at ion kinetic scales. Our simulations show two different behaviors depending on whether damping is present or not. Without Landau damping, modulational instability produces intense, filamentary magnetic structures with peak amplitudes reaching about 1.4 times the initial value. The resulting power spectrum follows $k_{\perp}^{-5/3}$ scaling in the inertial range ($k_{\perp}\rho_i < 1$), consistent with anisotropic MHD turbulence, and steepens to $k_{\perp}^{-8/3}$ at sub-ion scales. When Landau damping is included, the system loses 27.3\% of its initial magnetic energy over $100\,\omega_{ci}^{-1}$. The spatial distribution becomes much smoother as wave-particle resonances suppress small-scale fluctuations. The inertial range maintains $k_{\perp}^{-5/3}$ scaling, but the kinetic range steepens to $k_{\perp}^{-11/3}$, showing that energy is removed efficiently before the cascade reaches electron scales.

Comparison with MMS burst-mode observations supports this physical picture. Using Taylor's frozen-in hypothesis to relate frequency and wavenumber domains (Equation~\ref{eq:taylor}), the observed kinetic range slope of $f^{-3.33}$ lies between our undamped limit ($-2.67$) and strongly damped limit ($-3.67$). This suggests that magnetosheath turbulence operates in an intermediate damping regime where Landau dissipation is active but does not completely dominate the cascade. The spectral break near $f \approx 0.6$~Hz in the MMS data, corresponding to $k_{\perp}\rho_i \approx 1$, matches the transition scale found in our simulations. The presence of finite parallel electric fields (0.5--2~mV/m) and enhanced $|\delta E_{\perp}|/(|\delta B_{\perp}|v_A)$ ratios in the spacecraft data confirm the kinetic nature of the observed fluctuations, as expected from our model.

\begin{table}[h!]
\centering
\caption{Comparison of spectral properties between MMS observations and simulations. Under Taylor's frozen-in hypothesis, spectral indices are preserved between frequency and wavenumber domains, enabling direct comparison of slopes.}
\label{tab:spectral_comparison}
\begin{tabular}{l c c c}
\hline
\textbf{Spectral Property} & \textbf{MMS Observation} & \textbf{Undamped Sim.} & \textbf{Damped Sim.} \\
\hline
Inertial range index ($\alpha_1$) & $-1.97$ & $-1.67$ & $-1.67$ \\
Kinetic range index ($\alpha_2$) & $-3.33$ & $-2.67$ & $-3.67$ \\
Spectral break & $f \approx 0.6$~Hz & $k_{\perp}\rho_i \approx 1$ & $k_{\perp}\rho_i \approx 1$ \\
\hline
\multicolumn{4}{l}{\footnotesize Note: $f_{break} = 0.6$~Hz corresponds to $k_{\perp}\rho_i \approx 1$ for $V_{flow} \approx 430$~km/s (see Equation~\ref{eq:taylor}).} \\
\hline
\end{tabular}
\end{table}

The steeper inertial range observed by MMS ($f^{-1.97}$) compared to our simulations ($-5/3$) likely results from additional physical processes in the magnetosheath such as compressibility, intermittent structures, and effects of the bow shock that are not included in our incompressible two-fluid model. However, the agreement in the kinetic range, where KAW physics dominates, supports the conclusion that Landau damping is the primary mechanism for spectral steepening at sub-ion scales.

These results have implications for understanding energy dissipation in collisionless astrophysical plasmas. The fact that observed spectral slopes fall between our damped and undamped limits provides a framework for estimating effective damping rates from spectral measurements. Our findings also support the use of Landau-fluid approaches as an efficient alternative to full kinetic simulations for capturing wave-particle interaction physics. Future work extending this model to include temperature anisotropy and electron-scale dynamics would help to clarify how turbulent energy is partitioned between ions and electrons in the solar wind and planetary magnetosheaths.

\section{Data Availability}

The simulation code (modified NLSE solver written in Fortran 90), analysis routines, and Python plotting scripts used to generate the figures in this work are publicly available in the Zenodo repository at \url{https://doi.org/10.5281/zenodo.17847125}. The observational data from the Magnetospheric Multiscale (MMS) mission used in this study are publicly available from the NASA Coordinated Data Analysis Web (CDAWeb) database (\url{https://cdaweb.gsfc.nasa.gov/}).

\begin{acknowledgements}
The authors MKC, VS, and BS are grateful to the University Grants Commission (UGC), India, for providing financial support through the Non-NET Fellowship. The authors acknowledge the High Performance Computing (HPC) facility at Sikkim University for providing the computational resources used in this work.
\end{acknowledgements}

\bibliography{bibtex}{}
\bibliographystyle{aasjournalv7}

\end{document}